\documentstyle[aps,eqsecnum,psfig]{revtex}
\begin{document}

\draft

\preprint{\vbox{\hbox{DRAFT of U. of Iowa preprint 2000-2502}}}

\title{Examples of overlapping convergent expansions of scaling 
variables }

\author{Y. Meurice and S. Niermann\\ 
{\it Department of Physics and Astronomy, The University of Iowa, 
Iowa City, Iowa 52242, USA}}

\maketitle

\begin{abstract}

We construct series expansions for the scaling variables (which transform
multiplicatively under a renormalization group (RG) transformation) 
in examples where the 
RG flows, going from an unstable (Wilson's) 
fixed point to 
a stable (high-temperature) 
fixed point, can be calculated numerically. The examples are 
Dyson's hierarchical model and a simplified version of it.
We provide numerical evidence that the scaling variables about 
the two fixed points have overlapping domain of convergence.
We show how quantities such as the magnetic 
susceptibility can be expressed
in terms of these variables.
This procedure provide accurate analytical expressions 
both in the critical and high-temperature region.
\end{abstract}
\section{Introduction}

In many statistical mechanics or field theory problems 
one needs to  
calculate the  macroscopic 
features of a system in terms of some microscopic parameters.
In field theory language, the microscopic parameters are called
the bare parameters and the macroscopic behavior is encoded 
in the low momentum $n$-point functions which can be used to 
define the renormalized quantities. 
Expressing the renormalized quantities in terms of the
bare ones is a notoriously difficult problem.

The renormalization group (RG) method \cite{wilson71a,wilson74} was 
designed in part to tackle this problem. However, its practical 
implementation for realistic lattice models such as spin models or lattice
gauge theories is still a formidable technical enterprise.
Near RG fixed points, expansions are often available. The main problem
consists in interpolating between fixed points. 
An interesting attempt to model the interpolation between the weak and strong 
coupling regime of an asymptotically free gauge theory can be found 
in a recent publication of the QCD-TARO collaboration \cite{taro00}.
The interpolation was performed using the Monte Carlo simulation
following a method developed in Ref. \cite{gonzalez87}. 
One delicate step of this method is the choice of a finite number 
of operators which can be used to express the renormalized action 
after successive RG transformations.

In this article, we address the question of the interpolation 
between fixed points in models where the block spin method closes
exactly and where the 
RG transformation can be calculated easily by numerical methods.
One model satisfying these requirements is
Dyson's hierarchical model \cite{dyson69,baker72},
for which only the local measure is renormalized according to a rather
simple integral equation. This integral equation, briefly
reviewed in section \ref{sec:hm} can be treated 
with numerical integration  methods \cite{baker77,kim77}, or by
algebraic methods \cite{finite} based on the Fourier transform
of the original equation. 
The second method leads to very accurate determinations of the 
critical exponents \cite{gam3rapid,gam3}.

One could say that the hierarchical 
model (HM) is ``numerically solvable'', in the
sense that for a given set of bare parameters, one can calculate the 
zero-momentum $n$-point functions. However, one has to repeat the tedious
numerical
procedure for every particular calculation and 
analytical 
expressions are more desirable. 
In this article, we show that this goal can be achieved by 
constructing the scaling variables corresponding to 
the fixed points
governing the RG flows in the high-temperature (HT)
phase.

We use the terminology ``scaling variables''  for quantities which transform
multiplicatively under a RG transformation. In Ref. 
\cite{wegner72}, where these quantitites were introduced, 
Wegner call them ``scaling fields''. Indeed, in order to 
emphasize that they are functions of the bare parameters, we would 
prefer to call them
``scaling functions''. However since this term is already reserved for other 
quantities, we 
will use the terminology ``scaling variables'' which seems to have 
prevailed \cite{cardy96} over the years.

The scaling variables can 
in principle be constructed in the vicinity of any fixed point.
Near a given fixed point, they are simply the eigen-directions of the
linearized RG transformation. When moving away from the fixed point, 
non-linear terms need to be added order by order to maintain a 
multiplicative renormalization. What is the domain of convergence 
of this expansion? 
Can we combine two  expansions in order to follow the flows in
a crossover region?
These questions are in general very difficult to answer, but
if the domains of convergence of the expansions fail to overlap,
the whole approach is useless. The empirical 
results that we will present indicate that 
the domains of convergence of various
expansions do overlap. These results should be considered as 
an encouragement to pursue this approach for other models.

To limit the technical difficulties of constructing these functions
to orders large enough to get an idea about the asymptotic behavior of
their series expansions,
we have imposed 
some limitations on the calculations presented below. First, 
we only consider the HT phase. Second, we mostly consider the 
flows from the unstable fixed point to the HT fixed {\it starting
along the unstable direction}. The perturbative relaxation of this second 
condition will be briefly discussed. Third, we only discuss the magnetic 
susceptibility and not the higher moments.

The article is organized as follows. In section \ref{sec:hm}, we 
review the basic facts about the HM and its
RG transformation. We present a variant of the truncation
method proposed in Ref. \cite{finite} which will be used in the 
rest of the article. We also clarify the relationship between the 
truncation and the HT expansion. 

In the next four sections, we 
present, explain and illustrate the main ideas with a simplified 
one-variable model \cite{dual} 
where all the 
calculations are not too difficult. 
This model is simply a quadratic map with two fixed points, one stable
and one unstable.
The one-variable model is presented 
and motivated in section \ref{sec:toy}, the scaling variables are
constructed in section \ref{sec:scalingtm} and their convergence 
studied in section \ref{sec:largetm}. 
We want to make clear that all our assertions concerning the convergence
of series are based on the analysis of numerical values of the 
coefficients rather than on analytical results concerning these 
coefficients.
The susceptibility is calculated 
in section \ref{sec:sustm}. The most important result for these four 
sections is illustrated Fig. \ref{fig:tmoverlap} which shows that scaling
variables can be constructed accurately in overlapping regions.

The rest of the article is devoted to generalizing the construction
for the HM, for which the RG transformation can be approximated 
by a multivariable quadratic map. In section \ref{sec:multi}, we show
how to choose the coordinates in order to solve the linear problem.
In section \ref{sec:impl}, we present approximations which
allow one to calculate nonlinear expansions 
for the scaling variables. Finally, the questions of 
convergence are discussed in section \ref{sec:hmseries}. The most 
important result is illustrated in Fig. \ref{fig:hmoverlap} which
indicates overlapping domains of convergence.

In more general terms, our article addresses the question of the non-linear
behavior of the RG flows in models where non-linear expansions are 
calculable and can be compared with numerical solutions. In the first 
approximation, we have a linear theory. 
Trying to go beyond this first approximation has some similarities
with trying to make perturbation
near integrable systems in classical mechanics.

\section{DYSON'S HIERARCHICAL MODEL}
\label{sec:hm}

In this section, we review the basic facts about the RG transformation
of the HM to be used in the rest of the paper.
In order to avoid useless repetitions, we will refer the reader to 
Ref. \cite{finite} for a more complete discussion. In the following,
we will 
emphasize new material such as the 
various possibilities available for
the truncation procedure
and the relationship  between the truncation and the HT expansion.

\subsection{The RG transformation}

The energy density (or action in the field theory language) of the HM
has two parts. One part is non-local (the ``kinetic term'') and invariant 
under a RG transformation. Its explicit form can be found, for instance, 
in section II of Ref. \cite{hyper}. The other part is a sum of 
local potentials given in terms of a unique function $V(\phi)$. The
exponential ${\rm e}^{-V(\phi)}$ 
will be called the local measure and denoted $W_0(\phi)$.
For instance, for Landau-Ginsburg models, the measures are of the
form $W_{0}(\phi)= e^{-A \phi^2-B\phi^4}$, but we can also consider limiting 
cases such as 
a Ising measure $W_{0}(\phi)=\delta (\phi^2-1)$.  
Under a block spin transformation which integrates 
the spin variables in  ``boxes'' with two 
sites, keeping their sum constant, the local measure transforms according to 
the intergral formula
\begin{equation}
W_{n+1}(\phi) = \frac{C_{n+1}}{2} e^{(\beta/2)(c/4)^{n+1}\phi^2}
\int d \phi' W_{n} \left(\frac{ \phi - \phi'}{2} \right) W_{n} \left(
\frac{\phi + \phi'}{2} \right) \ , 
\end{equation}
where $\beta$ is the inverse temperature 
(or the coefficient in front of the kinetic term) 
and $C_{n+1}$ is a normalization
factor to be fixed at our convenience.

We use the Fourier transform
\begin{equation}
W_{n}(\phi) = \int \frac{d k}{2 \pi} e^{i k \phi} \hat{W}_{n}(k) \ .
\end{equation}
We introduce a rescaling of $k$ by a factor $u/s^{n}$, where $u$ and $s$ are
constants to be fixed at our convenience, by defining
\begin{equation}
R_{n}(k) \equiv \hat{W}_{n}(\frac{u k}{s^{n}}) \ ,
\end{equation}
In the following, we will use $s=2/\sqrt{c}$. For $c=2^{1-2/D}$, this 
corresponds to the scaling of a massless gaussian field in $D$ dimensions.
Contrarily to what we have done in the past, we will here absorb the 
temperature in the measure by setting $u=\sqrt{\beta}$.
With these choices, the RG transformation reads
%\begin{equation}
%R_{n}(k) \equiv \hat{W}_{n}\left[\frac{\sqrt{\beta} k}{(2/\sqrt{c})^{n}}
%\right]\ ,
%\end{equation}
%the recursion formula becomes
\begin{equation}
R_{n+1}(k) = C_{n+1} \exp \left[ -{1\over 2} 
{{\partial ^2} \over 
{\partial k ^2}} \right]\left[R_{n} \left({\sqrt{2}k\over 2} \right) \right]^2 \ . 
\label{eq:rec}\end{equation}
We fix the normalization constant $C_{n}$ so that $R_{n}(0) = 1$.
For an Ising measure, $R_{0}(k) = \cos(\sqrt{\beta}k)$, while in general,
we have to numerically integrate to determine the coefficients of
$R_{0}(k)$ expanded in terms of $k$.

If we Taylor expand about the origin,
\begin{equation}
R_{n}(k) = \sum_{l=0}^{\infty} a_{n,l} k^{2 l} \ ,
\end{equation}
(where $a_{n,0} = 1$) then the finite-volume susceptibility is
\begin{equation}
\chi_{n} = - 2 \frac{a_{n,1}}{\beta} \left(\frac{2}{c} \right)^{n} \ .
\label{eq:sus}
\end{equation}
The susceptibility $\chi$ is defined as
\begin{equation}
\chi \equiv \lim_{n \rightarrow \infty} \chi_{n} \ .
\end{equation}
The susceptibility tends to a finite limit for $0 \leq \beta < \beta_{c}$,
where $\beta_c$ is a constant depending on $c$.
For $\beta$ equal to or larger than
$\beta_{c}$, the definition of $\chi$ requires a subtraction 
(see e. g., Ref. \cite{hyper} for a practical implementation). 
In the following, we will only consider the 
HT phase ($\beta<\beta_c$).

We have eliminated $\beta$ from the mapping , moving it 
to the initial local measure.
The mapping has then fixed points independent of the
temperature.  
One of them is the ``universal function'' $U(k)$ found empirically in
Ref. \cite{gam3rapid,gam3} and which can be written with great
accuracy using numerical coefficients provided 
by Koch and Wittwer in Ref. \cite{wittwer95}.
The initial measure as a function of $\beta$ 
determines a line of
initial conditions
in the parameter space, running from the HT fixed
point, where all the initial parameters are zero,
to the critical surface ($\beta = \beta_{c}$).  

We can derive an explicit form for $a_{n+1,l}$
in terms of $a_{n,l}$.
\begin{equation}
a_{n+1, l} = \frac{u_{n,l}}{u_{n,0}} \ ,
\label{eq:aofu}
\end{equation}
where
\begin{equation}
u_{n,l} \equiv \sum_{i=0}^{\infty} \frac{(- \frac{1}{2})^{i} 
(2(l+i))!}{s^{2(l+i)} i! (2 l)!}
 \sum_{p+q=l+i} a_{n, p} a_{n, q} \ .
\label{eq:recursion}\end{equation}

%\begin{equation}
%a_{n+1, l} =  \frac{\sum_{i=0}^{\infty}
% B^i (2(l+i))!/(s^{2(l+i)} i! (2 l)!) \sum_{p+q=l+i}
% a_{n, p} a_{n, q}}
%{\sum_{i=0}^{\infty} B^i (2i)!/(s^{2 i} i!)
% \sum_{p+q=i} a_{n, p} a_{n, q}} \ .
%\label{eq:recursion}\end{equation}

\subsection{The high-temperature expansion}

\begin{figure}
\centerline{\psfig{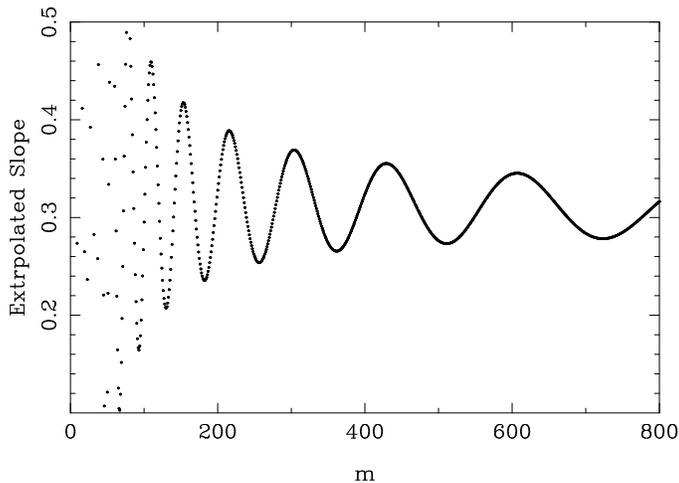}}
\caption{The extrapolated slope ($\hat{S}$) versus $m$ for 
the HM with $c=2^{1/3}$.}
\label{fig:hmexslope}
\end{figure}

To study the susceptibility not too far from the HT
fixed point, we can expand $\chi$ in terms of $\beta$.
We need to expand each of the parameters to some power in $\beta$ at each
level to find $\chi$. Since  we choose the scaling 
factor $u$ so that $\beta$ is eliminated from the recursion,
we find that $a_{0,l} \propto \beta^l$.  From the form of the
recursion, Eq.~(\ref{eq:recursion}), we can see that $a_{n,l}$
will always have $\beta^l$
as the leading power in its HT expansion (since $p+q\geq l$).  
If we want $R(k)$ expanded
to order $\beta^{m_{max}}$, we will 
use the truncated recursion formula
\begin{equation}
\left[ u_{n,l} \right]_{m_{max}}=\sum_{i=0}^{m_{max}-l} 
\frac{(-\frac{1}{2})^{i} (2(l+i))!}
{(4/c)^{(l+i)} i! (2 l)!} \sum_{p+q=l+i}\left[ a_{n, p} a_{n, q}\right]_
{m_{max}} \ ,
\label{eq:httruncation}
\end{equation}
where the notation $\left[\dots\right]_{m_{max}}$
means that the expression in brackets should be 
expanded up to order $m_{max}$ in $\beta$.
We define
the coefficients of the expansion of the infinite-volume susceptibility by
\begin{equation}
\chi(\beta) = \sum_{m=0}^{\infty} b_{m} \beta^m \ .
\end{equation}
We define $r_{m} \equiv b_{m}/b_{m-1}$, the ratio of two successive
coefficients, and introduce quantities \cite{nickel80},
called the extrapolated
ratio ($\hat{R}_{m}$) and the extrapolated slope ($\hat{S}_{m}$), defined
by
\begin{equation}
\hat{R}_{m} \equiv  m r_{m} - (m-1) r_{m-1} \ , 
\label{eq:rhat}
\end{equation}
and
\begin{equation}
\hat{S}_{m} \equiv  m S_{m} - (m-1) S_{m-1} \ , 
\label{eq:shat}
\end{equation}
where
\begin{equation}
S_{m} \equiv  \frac{-m(m-1) (r_{m} - r_{m-1})}{m r_{m} - (m-1)r_{m-1}}   \ , 
\end{equation}
is called the normalized slope.
If we use the expansion
\begin{equation}
\chi\simeq (\beta _c -\beta )^{-\gamma } (A_0 + A_1 (\beta _c -\beta)^{
\Delta }+\ldots )\ ,
\end{equation}
and assume that $A_{0}$ and $A_{1}$ are constants, we find
\begin{equation}
\hat{S}_{m} = \gamma - 1 - K m^{-\Delta} + O(m^{-2}) \ ,
\label{eq:estim}
\end{equation}
where $K$ is a constant.
However, if we calculate this quantity in $D=3$ with Ising and Landau-
Ginsburg measures, we find oscillations (Ref.~\cite{osc1,osc2}).
For comparison with results described later, Fig.~\ref{fig:hmexslope} shows 
these oscillations in the extrapolated slope.

\subsection{The truncation approximation}

The HT expansion can be calculated to very high order,
however, due to a large number of subleading corrections, 
this is a very inefficient way to obtain information 
about the critical behavior. In Ref. \cite{finite}, it was found that one 
can obtain much better results by altering Eq.  
(\ref{eq:httruncation}) .
First, we can retain a much smaller number of terms in the sum
(originally from 1 to $m_{max}-l$) than in the expansions 
$\left[\dots\right]_{m_{max}}$. As an example, one can calculate the 1000-th
HT coefficient of $\chi$ 
with 16 digits of accuracy using only 35 terms in the sum.
Second, we can simply replace the expansions $\left[\dots\right]_{m_{max}}$
by numerical values. This is equivalent to consider the polynomial
approximation
\begin{equation}
R_{n}(k) \simeq \sum_{l=0}^{l_{max}} a_{n,l} k^{2 l} \ ,
\end{equation}
for some integer $l_{max}$ at each $n$-step.
The mapping is then confined to an $l_{max}$-dimensional space.

There remains to decide if one should or not 
truncate to order $k^{2l_{max}}$
after squaring $R_n$. This makes a difference 
since the exponential of
the second derivative has terms with arbitrarily high order
derivatives. Numerically, one gets better results at intermediate
values of $l_{max}$ by keeping all the terms in $R_n^2$.
In addition, for the calculations performed later, the intermediate truncation 
pads the ``structure constants'' of the maps (see sec. \ref{sec:multi}) with
about fifty percent of zeroes. 
A closer look at section \ref{sec:multi}, may convince the reader that 
not truncating after 
squaring is more natural because we obtain 
correct (in the sense that they keep their 
value when $l_{max}$ is increased) structure constants in place of these 
zeroes. We have thus followed the second possibility where we truncate 
only once at the end of the calculation. With this choice
\begin{equation}
u_{n,l} \simeq \sum_{i=0}^{2l_{max}-l} \frac{(-\frac{1}{2})^{i} (2(l+i))!}
{(4/c)^{(l+i)} i! (2 l)!} \sum_{p+q=l+i} a_{n, p} a_{n, q} \ .
\label{eq:truncation2}\end{equation}

Compared to the HT expansion, the inital truncation to order $l_{max}$
is accurate up to order $\beta^{l_{max}}$. After one iteration, we 
will miss terms of order $\beta^{l_{max}+1}$ but we will also generate
some contributions of order $\beta^{2l_{max}}$. After $n$ iterations
we generate some of the 
terms of order $\beta^{2^nl_{max}}$ as in superconvergent 
expansions (such as Newton's method to calculate the roots
of a polynomial).

\section{A ONE-VARIABLE MODEL}
\label{sec:toy}

Before attacking the multivariable expansions of the scaling variables, we
would like to illustrate the main ideas 
and study the convergence of series with a simple one variable example
which retains the important features: a critical
temperature, RG flows going from an unstable fixed point to
a stable one, and log-periodic 
oscillations in the susceptibility.
%\subsection{The simplified recursion formula}

In order to obtain a simple one-variable model, we first consider the
$l_{max} = 1$ truncation using 
Eq. (\ref{eq:truncation2}).  
The
mapping is then reduced to only one variable.  The mapping takes
the form:
\begin{equation}
a_{n+1,1} = \frac{ (c/2)a_{n,1} -(3c^2/8)a_{n,1}^2}
{1 - (c/2)a_{n,1} + (3c^2/16)  a_{n,1}^2} \ ,
\end{equation}
Expanding the denominator and keeping terms
in the mapping only up to order $2$ in $a_{n+1,1}$, we obtain
\begin{equation}
a_{n+1,1} = (c/2) a_{n,1} -(c^2/8)a_{n,1}^2 \ .
\end{equation}
From Eq. (\ref{eq:sus}), we can put this recursion directly in terms
of the (truncation approximated) susceptibility:
\begin{equation}
\chi_{n+1} = \chi_{n} + \frac{\beta}{4} \left(\frac{c}{2} \right)^{n+1}
\chi_{n}^{2} \ .
\label{eq:tmchi}
\end{equation}

\begin{figure}
\centerline{\psfig{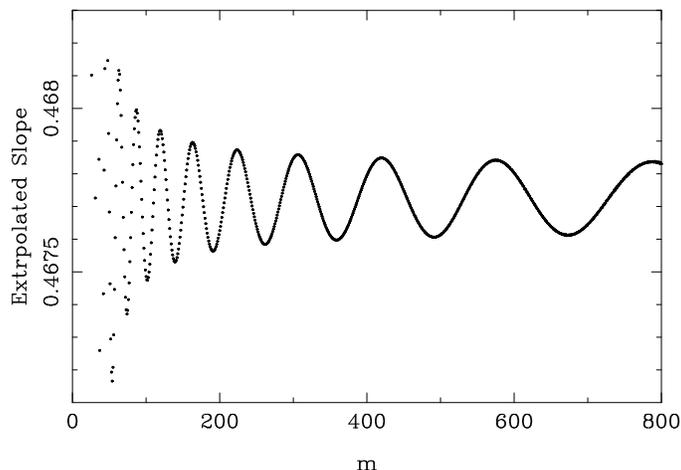}}
\caption{The extrapolated slope ($\hat{S}_m$) versus $m$ for the 
HT of $\chi$ calculated from the simplified recursion Eq. (\ref{eq:tmchi})
with $c=2^{1/3}$.}
\label{fig:tmexslope}
\end{figure}

This approximate equation was successfully  
used in Ref. \cite{finite} to model
the finite-size effects.
If we expand $\chi$ in $\beta$, and
define the exptrapolated slope, $\hat{S}_m$, as in Eq.~(\ref{eq:shat}),
we see oscillations in Fig.~\ref{fig:tmexslope}
quite similar to those in the HM.
(Fig.~\ref{fig:hmexslope}).

Our goal is to obtain the susceptibility (in the one-variable model)
as a function of $\beta$
and $c/2$.  Despite the simple appearance of the susceptibility recursion,
the behavior of the large-$n$ limit is not immediately apparent.
Simply running the recursion and examining
the result yields little, as the expression becomes very complicated
in short order.  Instead, we approximate the
susceptibility near to either $\beta = 0$ or $\beta = \beta_{c}$,
then successively build up more accurate
approximations as we move the temperature away from this point.

In the following, we use the notation $\xi \equiv c/2$ and we only consider 
the case $\chi_{0} = 1$ as in the Ising model.
Our analysis will be simplified if we remove the $n$-dependence
of the recursion.  For this purpose, we define a new quantity, 
$h_{n}$, such that
\begin{equation}
\chi_{n} \equiv \frac{\alpha h_{n}}{\xi^{n}} \ ,
\end{equation}
where $\alpha$ is an arbitrary constant.
This gives the recursion
\begin{equation}
h_{n+1} = \xi h_{n} + \frac{\beta}{4} \xi^{2} \alpha h_{n}^{2} \ .
\end{equation}
The fixed points of this map are
\begin{equation}
h^{\ast} = 0, \frac{4 (1-\xi)}{ \alpha \beta \xi^2} \ .
\end{equation}

We can choose $\alpha$ so that the non-zero fixed point is equal to one:
\begin{equation}
\alpha = \frac{4 (1-\xi)}{\beta \xi^2} \ .
\end{equation}
This has the nice effect of making the fixed points independent of
$\xi$ and $\beta$.  Also, $\beta$ is removed entirely
from the map, making it only dependent on $\xi$:
\begin{equation}
h_{n+1} = \xi h_{n} + (1-\xi) h_{n}^{2} \ .
\label{eq:hmap}\end{equation}
We call this map the ``$h$-map''.  We recover the susceptibility from:
\begin{equation}
\chi_{n} = \frac{4 (1-\xi)}{\beta \xi^{2}} \frac{h_{n}}{\xi^{n}} \ .
\end{equation}
Recalling that $\chi_{0} = 1$, we have
\begin{equation}
h_{0} = \frac{\beta \xi^{2}}{4 (1-\xi)} \ ,
\end{equation}
allowing us to write
\begin{equation}
\chi_{n} =  \frac{h_{n}}{h_{0}} \xi^{-n} \ 
\end{equation}
for non-zero $h_{0}$ ($h_{0} = 0$ means that $\beta = 0$ and 
$\chi_{n} = 1$ for all $n$).

\begin{figure}
\centerline{\psfig{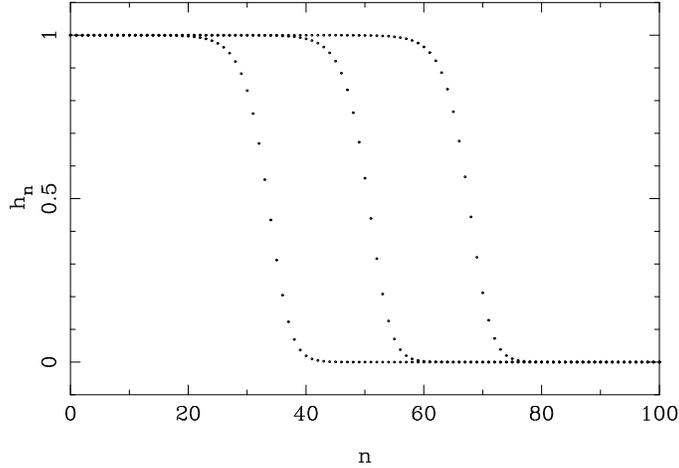}}
\caption{$h_{n}$ versus $n$ for $\xi = 0.5$.  Here $h_{0} = 1 - 10^{-6}$,
 $1 - 10^{-9}$, and $1 - 10^{-12}$.}
\label{fig:h5}
\end{figure}

If we iterate the recursion Eq. (\ref{eq:hmap}), we find that
for $0 \leq h_{0}<1$,
$h_n \rightarrow 0$ as $n \rightarrow \infty$.  Correspondingly,
for $n \rightarrow \infty$, $\chi_{n}$ approaches a
constant, $\chi$, as we saw by using the original recursion.
For $h_{0} = 1$, $h_{n} = 1$ for all $n$.  For $h_{0} > 1$,
$h_n \rightarrow \infty$ for $n \rightarrow \infty$.
This is the same behavior of the susceptibility expected as the temperature
($\beta$) crosses the critical value.
The adjustable parameter in $h_{0}$ is
$\beta$, so we see that $h_{0} = 1$ corresponds to $\beta = \beta_{c}$,
consequently
\begin{equation}
\beta_{c} = \frac{4 (1-\xi)}{\xi^{2}} \ .
\end{equation}
In Fig .~\ref{fig:h5}, we show the values of $h_n$ for different
values of $h_0$. This figure indicates that the shape of the function
is independent of $h_0$.  
\newpage
%The smaller $\xi$ is,
%the more quickly
%$h_{n}$ drops from its initial value.  Thus for $\xi = 0.1$, we use $h_{0}$
%very close to $1$.
%The shape of the transition region, from $h_{n}$ near $1$
%down to $h_{n}$ near $0$, also depends on the value of $\xi$.
%For smaller $\xi$,
%the curve drops more quickly than for $\xi$ near to $1$.  Also the 
%curve becomes more symmetric upon a ``diagonal flip'' about the center of the
%transition region, as $\xi$ approaches $1$.

To understand the behavior of the $h$-map, we need to investigate
the stability of the fixed points.  We find:
\begin{equation}
\left(\frac{d h_{n+1}}{d h_{n}} \right)_{h_{n}=0} = \xi \ , 
\end{equation} and
\begin{equation}
\left(\frac{d h_{n+1}}{d h_{n}} \right)_{h_{n}=1} = 2 - \xi \ .
\end{equation}
In the following, we only consider $0\leq\xi\leq1$, so the zero fixed
point is stable, while the other is unstable.  
All the $h_{0}$ between zero and the unstable fixed point are attracted toward
the stable fixed point.
Also note that for all $h_{0}$ greater than the unstable fixed point,
$h_{n} \rightarrow \infty$.
For $h_{0}$ at the unstable fixed point, $h_{n} = 1$ for all $n$. So
then $\chi_{n} = \xi^{-n}$.  This diverges,
since $\xi<1$.

Note that the $h$-map can be put in the form 
used in Collet and Eckmann monography \cite{collet80},
\begin{equation}
x_{n+1}=1-\mu x_n^2\ ,
\end{equation}
with 
\begin{equation}
\mu=(\xi/4)(\xi-2) \ ,
\end{equation}
by using a linear transformation.
With their parametrization, the first bifurcation where the fixed point
becomes unstable and a cycle 2 develops occurs at $\mu =0.75$. This 
corresponds to $\xi$=3 or -1 which is outside of the region where 
we will study the map in the following
(clearly, a cyclic behavior means no thermodynamics limit).
%\section{The $d$-map}

We can expand the map about the unstable fixed point, $h_{n} = 1$.
\begin{equation}
h_{n+1} = 1 + (2-\xi)(h_{n}-1) + (1-\xi)(h_{n}-1)^{2} \ .
\end{equation}
The usual notation for the eigenvalue near the critical point is
$\lambda$, so we have $\lambda \equiv 2-\xi$.  If 
we define  
\begin{equation}
d_{n} \equiv 1-h_{n}\ ,
\label{eq:doh}
\end{equation}
then:
\begin{equation}
d_{n+1} = \lambda d_{n} + (1-\lambda) d_{n}^{2} \ ,
\label{eq:dmap}\end{equation}
with the starting value $d_{0} = 1-\beta/\beta_{c}$.
We call this map the ``$d$-map''.  

Note the similarity of the $d$-map to the
original $h$-map.  We can introduce a duality transformation \cite{dual} 
between
the two maps which interchanges
$h_{n} \leftrightarrow d_{n}$ and $\xi  \leftrightarrow \lambda$.
If the duality transformation
is applied twice, we return to the original quantities.
For $0<h_0<1$, we also have $0<d_0<1$ with small values 
(approaching 0 from above)
in one variable corresponding to large values (approaching 1 
from below) values in the dual variable.
%\section{The linear approximation}

We would like to construct the susceptibility, $\chi$, as
a function of $d_0$.  As a first approximation, we use
Eq.~(\ref{eq:dmap}) to linearize near the unstable fixed point
($d_{0} = 0$), finding the critical exponent $\gamma$ in the process.
Beginning with a value of $d_{0}$ very small ($\beta$ close to
$\beta_{c}$), then for a certain number of iterations $d_{n+1} \simeq
\lambda d_{n}$, so that $d_{n} \simeq \lambda^{n} d_{0}$.  So as
long as $\lambda^{n} (1-\beta/\beta_{c}) << 1$, then
$h_{n} \simeq 1$ and 
$\chi_{n} \simeq \xi^{-n}$.  If we assume that $h_{n}$ stays near
1 for some number of iterations, and then drops quickly, near some
$n=n^{\ast}$, to the region where $h_{n} \propto \xi^{-n}$ (thus
stabilizing $\chi$), then
$\chi \sim \xi^{-n^{\ast}}$.

Defining $n^{\ast}$ by $\lambda^{n^{\ast}} d_{0} = 1$, we have
\begin{equation}
n^{\ast} = -\frac{\ln d_{0}}{\ln \lambda} \ ,
\end{equation}
which gives
\begin{equation}
\chi \sim \xi^{-n^{\ast}} =
(1-\frac{\beta}{\beta_{c}})^{\frac{\ln \xi}{\ln \lambda}} \ .  
\end{equation}
Using the usual notation for the critical exponent, the leading
singularity is given by $(1-\beta/\beta_{c})^{-\gamma}$.  So we have
\begin{equation}
\gamma = \frac{\ln(1/\xi)}{\ln \lambda} \ ,
\end{equation}
or, equivalently:
\begin{equation}
\lambda^{\gamma} = \frac{1}{\xi} \ .
\end{equation}

\begin{figure}
\centerline{\psfig{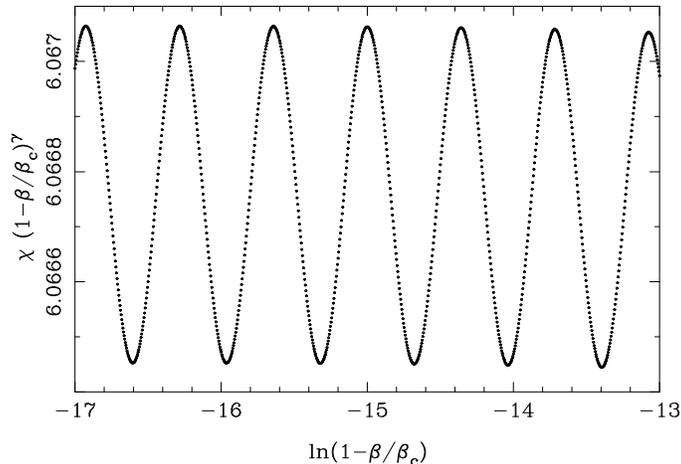}}
\caption{Oscillations near the critical point.  The susceptibility
is generated from the recursion Eq. (\ref{eq:dmap}) with 
$\lambda = 1.9$, while $\gamma$ and $\beta_{c}$ are calculated
from the formulas derived in the text. }
\label{lbet}
\end{figure}

We can divide the leading singularity out of $\chi$ 
and plot the remainder near to the critical point
(Fig.~\ref{lbet}).
\noindent
We see 
periodic oscillations with respect to 
the variable $\ln (1-\frac{\beta}{\beta_{c}})$ which have period of
$\ln \lambda$.  The linear approximation gives no clue as to the
origins of the oscillations.  To understand this, as well as the
higher-order corrections to the leading singularity, we need to expand
$d_{n}$ to higher order in $d_{0}$.
\section{SCALING VARIABLES IN THE ONE-VARIABLE MODEL}
\label{sec:scalingtm}

In this section, we show that the idea of scaling variable comes naturally
as a way to express the susceptibility of the one-variable model 
in terms of the 
input parameters.

\subsection{$d_n$ as a function of $d_0$}

We can find $d_{n}$ in terms of $d_{0}$ by expanding
\begin{equation}
d_{n} = \sum_{l=1}^{\infty} q_{n,l} \, d_{0}^{l} \ .
\end{equation}
and plugging into Eq. (\ref{eq:rec}). We find
\begin{equation}
q_{n,l} = \lambda q_{n-1,l} + (1-\lambda)
 \sum_{j=1}^{l-1} q_{n-1,j} \, q_{n-1,l-j} \ .
\end{equation} 
Given the initial coefficients, i.e. $q_{1,1} = \lambda$,
 $q_{1,2} = (1-\lambda)$, and $q_{1,j} = 0$ for $j>2$, we can find all
the coefficients at any $n$.  For example,
\begin{equation}
q_{n,2} = \frac{\lambda^{n} (1-\lambda^n)}{\lambda} \ .
\end{equation}
Finding forms for the coefficients explicitly in terms of $\lambda$ and
$n$ by this direct method quickly becomes difficult. We will show that
this task can be simplified by introducing the scaling variables.

As noticed before, the transition region
from $d_{n} \simeq 0$ to $d_{n} \simeq 1$ has a shape 
which looks independent
of $d_{0}$ (see Fig.~\ref{fig:h5} and use Eq. (\ref{eq:doh})).
The shape seem to depend only on the value of $\lambda$. We can prove this 
in a restricted way by imagining a new initial value of ``$d_{0}$'' which
is identical to the current $d_{1}$.  The shape of the 
curve to the right would
then be identical to the current case.  Likewise, we can reverse the
map, getting new values for ``$d_{0}$''
which nonetheless generate all the values in the current series
$d_{0},d_{1}, \ldots$. This inverse map is unique, assuming we
confine it to only positive values for $d_{n}$, and has the form
\begin{equation}
d_{n} = \frac{\sqrt{\lambda^2+4(1-\lambda)d_{n+1}} -
 \lambda}{2(1-\lambda)} \ .
\end{equation}
In this case, the renormalization group is a group in the strict
sense, since we have defined a unique inverse.

What we would like then is a way of parameterizing $d_{n}$ 
in terms of a function independent of $n$.
Regardless of our actual $d_{0}$, we can extrapolate backwards, as
suggested above, to another ``$d_{0}$'', which is small enough so that
the linearized method works, where each new ``$d_{n}$'' scales
approximately as $\lambda^{n}$ times the ``$d_{0}$''.  This suggests
using a new parameter, which we call $y_{n}$, that scales exactly
like this, so that if $d_{n}$ corresponds to $y_{n}$,
then $d_{n+1}$ corresponds to $y_{n+1} = \lambda y_{n}$.  This is
our first covariant quantity, so-called since it scales exactly
the same, regardless of the value of $n$.

The idea of using variables with simple transformation
properties has a long history, for instance the angle-action variables in 
classical mechanics and the normal form of differential equations
appearing in Poincar\'e's dissertation \cite{arnold88}.
For continuous RG transformations, Wegner\cite{wegner72} introduced
the notion of ``scaling field''. We discuss here its analog for 
discrete RG transformations.

\begin{figure}
\centerline{\psfig{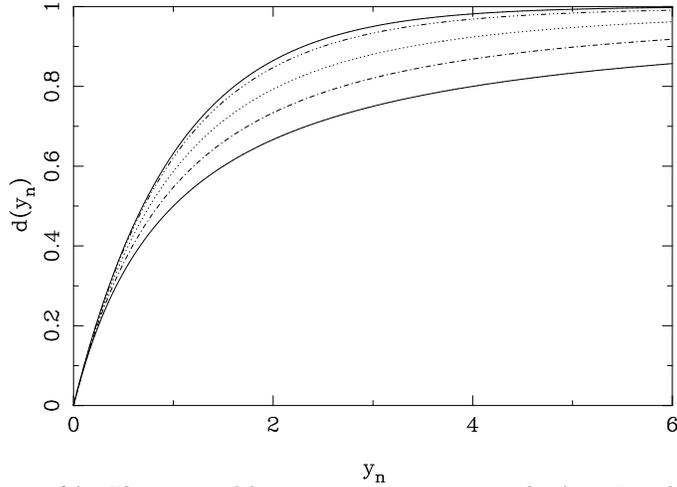}}
\caption{$d(y_{n}) $ for various values of $\lambda$.  The top and
bottom curves represent the $\lambda \rightarrow 2$ and $\lambda
\rightarrow 1$ limits, respectively.  The dash-dot-dot line is
$\lambda = 1.9$, the dotted line is $\lambda = 1.6$, and the
dash-dot line is $\lambda = 1.3$.}
\label{dy}
\end{figure}

Let us define a function $d$ such that $d(y_{n}) \equiv d_{n}$.
From 
the explicit form of the recursion  
formula Eq. (\ref{eq:rec}) this requirement implies 
\begin{equation}
d(\lambda y_{n}) = \lambda d(y_{n}) + (1-\lambda) d^{2}(y_{n}) \ .
\end{equation}
If we let
\begin{equation}
d(y_{n}) = \sum_{l=1}^{\infty} s_{l} \, y_{n}^{l} \ ,
\end{equation}
we find that
\begin{equation}
s_{l} = \frac{1-\lambda}{\lambda^{l}-\lambda} \sum_{j=1}^{l-1} s_{j} \, s_{l-j} \ .
\label{eq:coef}\end{equation}
The first coefficient is undetermined.  We let $s_{1} \equiv 1$, so that
for small $y_{n}$, we have $d_{n} \simeq y_{n}$.  The
 first few coefficients give
\begin{equation}
d(y_{n}) = y_{n} - \frac{1}{\lambda} y_{n}^{2} +
 \frac{2}{\lambda^{2}(\lambda +1)} y_{n}^{3}+\ldots
\end{equation}
In Fig.~\ref{dy}, we plot $d(y_{n})$ for several values of $\lambda$.

%Below, we find the formulas for $\lambda$ approaching  $1$ and $2$. 
%While $n$ can have arbitrarily large negative values (where the inverse map
%is allowed), $y_{n}$ can have a value no smaller than $0$.  For any
%given initial value for $d_{0}$, $n \rightarrow - \infty$ corresponds
%to $y_{n} \rightarrow 0$.  Since $\lambda > 1$, we have
%$n \rightarrow \infty$ corresponding to $y_{n} \rightarrow \infty$.
In the limit
where $\lambda \rightarrow 2$, we can calculate explicitly  $d(y_{n})$.
This corresponds to $\xi = 0$, so
that
\begin{equation}
h_{n+1} = h_{n}^{2} \ .
\end{equation}
This is satisfied if $h_{n} = e^{a y_{n}}$, where $a$ is arbitrary.  So
we have $d_{n} = 1 -  e^{a y_{n}}$.
To satisfy $y_{n} \simeq d_{n}$ for small $d_{n}$, we need $a = - 1$.
Finally, we have
\begin{equation}
d(y_{n}) = 1 -  e^{-y_{n}}\ .
\end{equation}

\begin{figure}
\centerline{\psfig{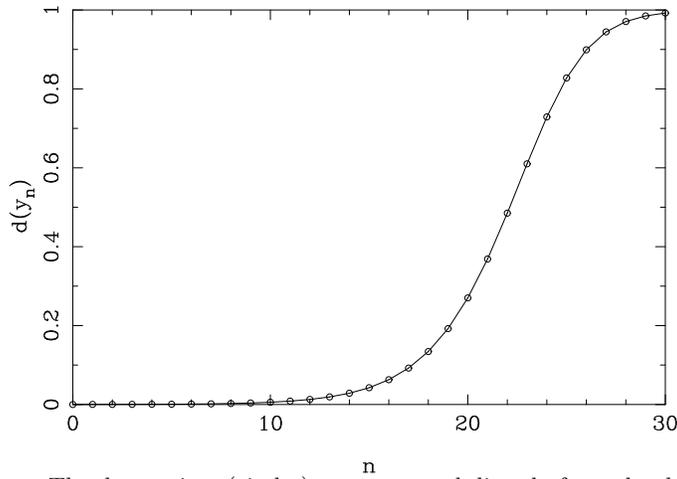}}
\caption{$d(y_{n})$ plotted against $n$.  The data points (circles)
are generated directly from the $d$-map recursion with $\lambda = 1.5$ 
and $d_{0} = 10^{-4}$.}
\vskip0.2pt
\label{dn}
\end{figure}

The expression of $d_n$ as $d(\lambda^n y_0)$ allows us to interpolate 
between integral values of $n$.
In Fig.~\ref{dn}, we show the curve for $d(y_{n})$
superimposed on points generated from the $d$-map for $\lambda = 1.5$.
\noindent
We can invert the series for $d(y_{n})$ in terms of $y_{n}$,
defining a new
function $y$, so that $y(d_{n}) = y_{n}$.
Since we know the coefficients of the $d(y_{n})$ series,
we can invert to get $y$ in terms of $d_{n}$.
More directly, we can use $y_{n+1} = \lambda y_{n}$, or 
\begin{equation}
y(\lambda d_{n} + (1-\lambda) d_{n}^{2}) = \lambda y(d_{n}) \ ,
\end{equation}
along with the constraint $y(d_{n}) \simeq d_{n}$ for small values of $d_{n}$,
to construct a series solution.
If we let
\begin{equation}
y(d_{n}) = \sum_{l=1}^{\infty} t_{l} \, d_{n}^{l} \ ,
\end{equation}
we find that
\begin{equation}
t_{l} = \frac{-1}{\lambda^{l}-\lambda} \sum_{j=1}^{int(l/2)}
\frac{(l-j)!}{j! (l-2j)!}  \lambda^{l-2j} (1-\lambda)^{j} t_{l-j} \ ,
\end{equation}
with $t_{l} \equiv 1$.  The notation $int(l/2)$ means to use the integer
part of $l/2$.
The first few terms give:
\begin{equation}
y(d_{n}) = d_{n} + \frac{1}{\lambda} d_{n}^{2} +
 \frac{2}{\lambda(\lambda +1)} d_{n}^{3}+\ldots
\end{equation}
Taking our $\lambda \rightarrow 2$ example, we can immediately invert
our result for $d(y_{n})$, to find
\begin{equation}
y(d_{n}) = - \ln (1-d_{n}) \ .
\end{equation}
With the $y$-function, we can now reconstruct $d_{n}$ in terms of $d_{0}$:
\begin{equation}
d_{n}=y^{-1}(\lambda^ny(d_0)) \ .
\label{eq:dsubn}
\end{equation}

\subsection{$h_{n}$ as a function of $h_{0}$}

Because of the duality between our two maps, we can
easily reproduce all of the above results of the $d$-map for
the $h$-map.  Just as with $d_{n}$, we can parameterize $h_{n}$ in
terms of a covariant variable, $\tilde{y}_{n}$, dual to $y_{n}$.
Similarly to before, we define $h(\tilde{y}_{n}) \equiv h_{n}$, with
$\tilde{y}_{n} = \xi^{n} \tilde{y}_{0}$.  We can also define the inverse
function for $h$, $\tilde{y}$, where
\begin{equation}
\tilde{y}(\xi h_{n} + (1-\xi) h_{n}^{2}) = \xi \tilde{y}(h_{n}) \ .
\end{equation}
We can immediately carry over all of the results we found for $d(y_{n})$
and $y(d_{n})$ to $h(\tilde{y}_{n})$ and $\tilde{y}(h_{n}$), replacing 
$\lambda$ with $\xi$.  For example,
\begin{equation}
h(\tilde{y}_{n}) = \tilde{y}_{n} - \frac{1}{\xi} \tilde{y}_{n}^{2}
 + \frac{2}{\xi^{2}(\xi +1)} \tilde{y}_{n}^{3}+\ldots \ ,
\label{eq:hofytilde}\end{equation}
while
\begin{equation}
\tilde{y}(h_{n}) = h_{n} + \frac{1}{\xi} h_{n}^{2} +
 \frac{2}{\xi(\xi +1)} h_{n}^{3}+\ldots
\end{equation}
As with $d_{n}$, we can find $h_{n}$ to any order in $h_{0}$.
The first few terms in the expansion give:
\begin{equation}
h_{n} = \xi^{n} h_{0}+ \frac{\xi^{n}-\xi^{2 n}}{\xi} h_{0}^{2} + \ldots
\end{equation}

The new covariant variable behaves similarly to $y_{n}$.  However,
since $\xi < 1$, we see that as $n \rightarrow \infty$, $\tilde{y}_{n}
\rightarrow 0$, while as $n \rightarrow -\infty$, $\tilde{y}_{n}
\rightarrow \infty$.  As with the $d$-map, $\tilde{y}_{n}$ goes to
zero with $h_{n}$, while blowing up as $h_{n}$ approaches $1$.  Thus, for
various $\xi$,
$h$ versus $\tilde{y}_{n}$ and $\tilde{y}$ versus $h_{n}$ will have
at least qualitatively similar shapes to the corresponding $d$-map plots.

\section{Large order behavior for the one-variable model}
\label{sec:largetm}

In this section we give empirical results concerning the large order
behavior of $y(d)$, $\tilde{y}(h)$ and their inverses. 
From this, we infer the domain of convergence of the expansions.
We would like to know if the series converge
for the ranges of interest.  The allowed ranges are $0$ to infinity
for $y_{n}$ and $\tilde{y}_{n}$, and $0$ to $1$ for $d_{n}$ and $h_{n}$.

\subsection{$d(y)$}

For the $d$-map, we first examine the extreme allowed values for
$\lambda$, $1$ and $2$.
For $\lambda = 2$ ($\xi=0$), we already found 
that $d(y) = 1-e^{-y}$.  Eq. (\ref{eq:coef}) 
converges in the whole complex plane.
At the other end of our range, for $\lambda \rightarrow 1$
($\xi \rightarrow 1$), Eq. (\ref{eq:coef}) becomes
\begin{equation}
s_{l} = \frac{-1}{l-1} \sum_{j=1}^{l-1} s_{j} s_{l-j} \ ,
\end{equation}
so that
\begin{equation}
d(y) = y- y^{2} + y^{3} - y^{4} + \ldots
\end{equation}
This converges only for $0<y<1$, to $d(y) = y/(1+y)$.

\begin{figure}
\centerline{\psfig{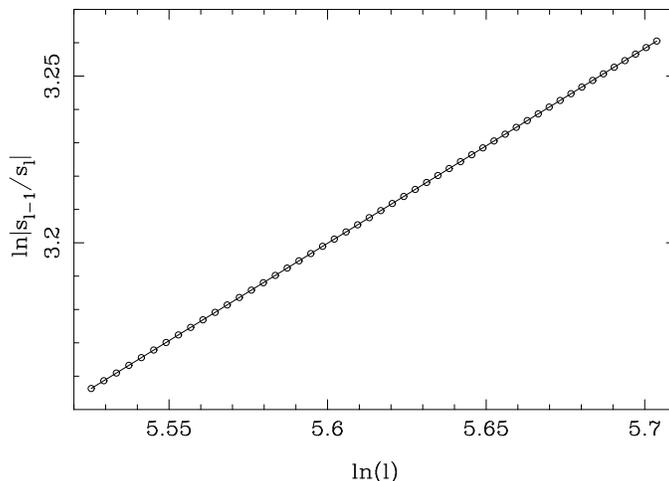}}
\caption{Linear fit to a logarithmic plot of the ratio of coefficients
against the order of the bottom coefficient.  Here, $\lambda = 1.5$.
The slope of the line is 0.584. }
\label{line}
\vskip0.2pt
\end{figure}

\begin{figure}
\centerline{\psfig{figure=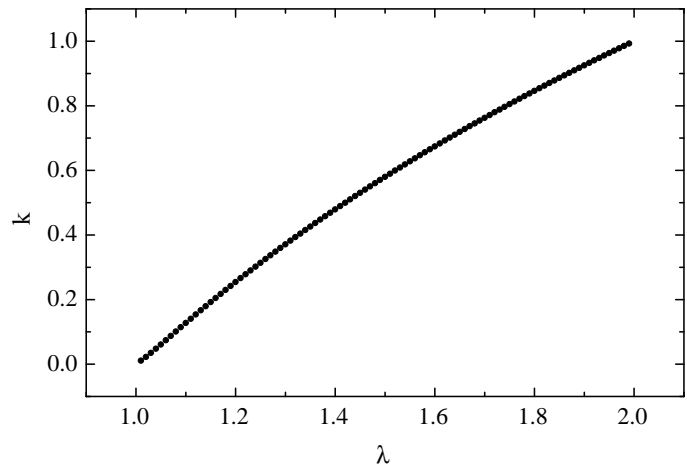,width=4in}}
\caption{The exponent $k$ defined in Eq. (\ref{eq:kexp}) as a function 
of $\lambda$. }
\label{fig:slope}
\end{figure}

We use the ratio test to determine convergence for our series.
The series will converge 
in the entire complex plane if $|s_{l-1}/s_{l}|$ increases without
bound as $l \rightarrow \infty$.
For all tested values of $1<\lambda<2$, we found that we could fit
a straight line to plots of $\ln |s_{l-1}/s_{l}|$ versus $\ln (l)$ for large
enough $l$ (Fig.~\ref{line}).
Thus for large $l$, the coefficients follow the rule
\begin{equation}
|\frac{s_{l}}{s_{l+1}}| \simeq C l^{k} \ ,
\label{eq:kexp}
\end{equation}
where we find that $C$ is always of order $1$ (in fact, $0.9<C<1$) and $0<k<1$.
The value of $k$ can be determined 
by a linear fit.
It seems clear that with $1<\lambda<2$, the series for $d(y)$ converges
for all $0<y_{n}<\infty$.  In agreement with the above exact
series, for $\lambda \rightarrow 1$, $C \rightarrow 1$ and $k
\rightarrow 0$, while for $\lambda \rightarrow 2$, $C \rightarrow 1$
and $k \rightarrow 1$.
The intermediate values are given in Fig. \ref{fig:slope}.

The form that the ratio of successive coefficients
takes also gives us an idea of the number of coefficients we will
need to approximate $d(y)$ for a given value of $y$.  Assuming that
a particular term in the series becomes significant when it is the
same size as the previous term, we find that the maximum order that
we need in our approximation goes like
\begin{equation}
l \sim y^{1/k} \ .
\end{equation}
Thus for $\lambda$ near $2$, the number of terms needed
will grow approximately linearly with $y$.  On the other hand,
for $\lambda$ near 1, where $k \rightarrow 0$, the number of terms
will increase much more quickly with $y$.

\subsection{$h(\tilde{y})$}
\begin{figure}
\vskip0.2in
\centerline{\psfig{figure=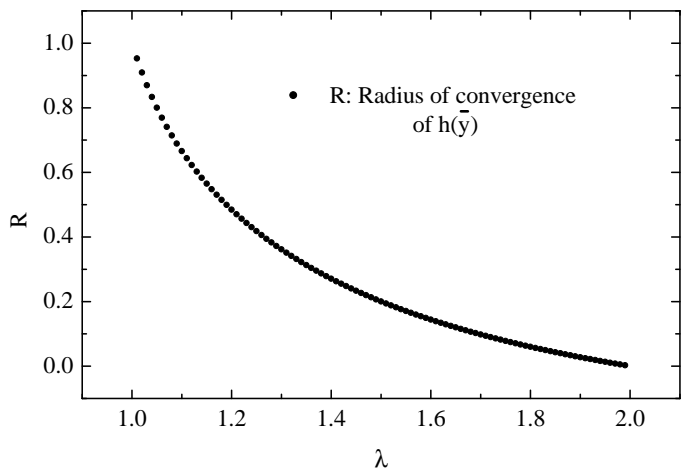,width=4in}}
\caption{Radius of convergence of $h(\tilde{y})$ as a function of 
$\lambda$. }
\label{fig:radius}
\end{figure}

We now turn to the dual function, $h$.  Above, we found
the series in $d(y)$ for $\lambda \rightarrow 1$.
From this, we can immediately write down $h(\tilde{y})$ when $\xi
\rightarrow 1$.  The series is identical in form, with
$\lambda$ replaced by $\xi$,  so that
\begin{equation}
h(\tilde{y}) = \tilde{y} - \tilde{y}^{2} + \tilde{y}^{3}
+ \ldots \ ,
\end{equation}
which converges for $0<\tilde{y}<1$ to $h(\tilde{y}) = 
\tilde{y}/(1+\tilde{y})$.
For the general case, we again examine the ratios of successive
coefficients, $|s_{l}/s_{l+1}|$.
We find the ratios flatten to constant values, for large enough
$l$ ,
%(Fig.~\ref{hco})
indicating finite radii
of convergence.  The radii get smaller
and vanish as $\xi$ approaches zero as shown in Fig. \ref{fig:radius}.
Note that for any value of $\xi$  tried, we found good evidence that
$lim_{n\rightarrow\infty}\widehat{S}_n=-1.5$, indicating a $(\tilde{y}-
\tilde{y}_c)^{1/2}$ behavior. This is consistent with the existence of
a quadratic minimum for the inverse function discussed below.

\subsection{$y(d)$)}

We already found above that for $\lambda=2$, $y(d)= -\ln(1-d)$.
This series converges for $0<d<1$, which is the required range.
(Since $y$ must diverge as $d\rightarrow 1$, it is not
surprising that our series breaks down at this point.)
For $\lambda \rightarrow 1$,
we have, inverting the $d$-function found above, that
$y(d) = d/(1-d)$
which also converges for $0<d<1$.
We find empirically from the analysis of ratios 
that for all $1<\lambda<2$, 
$y(d)$ converges in the region $0<d<1$.

The analysis of the extrapolated slope for various $\lambda$ gives 
convincing evidence that the main singularity has the form 
\begin{equation}
y(d)\sim d^{-1/\gamma}\ . 
\end{equation}
This can be seen with short series when 
$\lambda$ is close to one and requires larger and larger series as 
$\lambda$ gets close to 2.
\subsection{$\tilde{y}(h)$)}

\begin{figure}
\centerline{\psfig{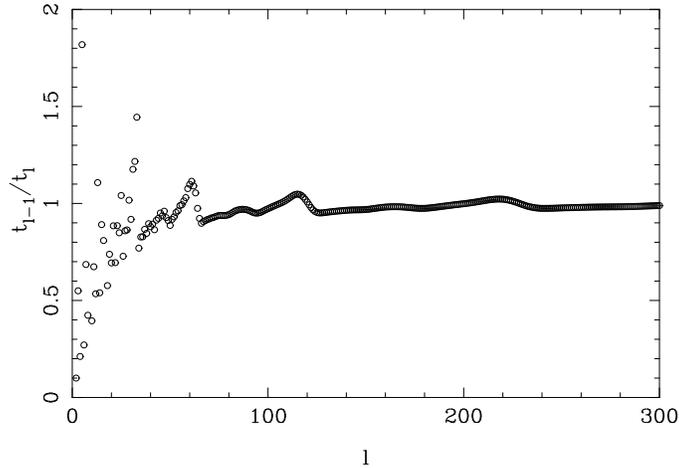}}
\caption{Ratios of coefficients for $\tilde{y}(h_{n})$ with $\xi = 0.1$.}
\label{ytco2}
\end{figure}

The $\tilde{y}(h)$ series behaves
similarly to the $y(d)$, having a radius
of convergence of $1$ for all $\xi$.
In this case, the ratio of coefficients $t_{l}/t_{l+1}$ approaches $1$
smoothly from below as $l$ increases, for larger $\xi$ 
%(Fig.~\ref{ytco}).
Around $\xi = 0.4$, small oscillations begin to appear in the curve.
As $\xi$ becomes smaller, these oscillations grow and become quite
large for $\xi$ smaller than $0.1$ (Fig.~\ref{ytco2}).
However, in all cases studied, the ratio
eventually approaches $1$, for large enough $l$.

A detailed study shows that if we continuate $\tilde{y}(h)$ for negative 
values of $h$ using the series expansion, the function develops a quadratic 
minimum when $h\rightarrow -1$. The absolute value of this minimum found 
in all examples studied is compatible with the radius of convergence
of the inverse function and the square root singularity of the 
$h(\tilde{y})$ discussed above.

As one can guess by looking at Fig. \ref{fig:tmexslope}, the 
analysis of the extrapolated slope 
is intricated. However, if we calculate enough terms and if $\lambda$ 
is not too close to 2, one can get approximate results which
are consistent with a main singularity of the form 
\begin{equation}
\tilde{y}(h)\sim h^{-\gamma}\ .
\end{equation}
For instance, just by looking at 
the asymptotic behavior of 
Fig. \ref{fig:tmexslope}, one can see that $\gamma \simeq 1.4677$, 
as expected, with errors of the order $10^{-4}$.

\section{The susceptibility of the one-variable model}
\label{sec:sustm}

Now that we can calculate $h_{n}$ and $d_{n}$, we can construct the
(infinite-volume limit) susceptibility.  If we focus on the HT
fixed point and the $h$-map, we get the HT expansion in
terms of $h_{0} = \beta/\beta_{c}$.  On the other hand, to understand
the susceptibility's behavior near to the fixed point, we need to expand
in terms of $d_{0} = 1 - \beta/\beta_{c}$. Finally, we consider the
possibility of combining the two expansions in the crossover region.

\subsection{The HT expansion of the susceptibility}
\label{susceptibility}

We could find the HT expansion from the finite-volume susceptibility
found in the previous section, by taking the large-$n$ limit.  Since
$0<\xi<1$, all of the $\xi^{n}$ terms will drop out.  However, we
can also construct the large-$n$ limit directly.
Recalling that $\chi_{n} = (h_{n}/h_{0}) \xi^{-n}$, and using
$\tilde{y}_{n} = \xi^{n} \tilde{y}_{0}$, we find that
\begin{equation}
\chi_{n} = \frac{h_{n} \tilde{y}_{0}}{h_{0} \tilde{y}_{n}}
= \frac{h(\tilde{y}_{n}) \tilde{y}_{0}}{h_{0} \tilde{y}_{n}} \ .
\end{equation}
The infinite-volume limit becomes:
\begin{equation}
\chi \equiv \lim_{n \rightarrow \infty} \chi_{n} = 
\frac{\tilde{y}_{0}}{h_{0}} \lim_{n \rightarrow \infty} \frac{h(\tilde{y}_{n})}
{\tilde{y}_{n}} \ .
\end{equation}
Using the $\tilde{y}$-expansion for $h$, we find
\begin{equation}
\frac{h(\tilde{y}_{n})}{\tilde{y}_{n}} = 1 - \frac{\tilde{y}_{n}}{\xi} + \ldots
\label{eq:hz}\end{equation}
As $n \rightarrow \infty$, $\tilde{y}_{n} \rightarrow 0$,
so the infinite volume limit of the susceptibility is:
\begin{equation}
\chi = \frac{\tilde{y}_{0}}{h_{0}} = \frac{\tilde{y}(h_{0})}{h_{0}} \ .
\label{eq:suscep}\end{equation}
Using the $h$-expansion for $\tilde{y}$, we get the HT
expansion for the susceptibility:
\begin{equation}
\chi = 1 + \frac{1}{\xi} h_{0} + \frac{2}{\xi(1 + \xi)} h_{0}^2 + \ldots
\end{equation}

Using $\chi_{n} = (h_{n}/h_{0}) \xi^{-n}$, we find the HT
expansion for the finite-volume susceptibility (recall that $h_{0} = 
\beta/\beta_{c}$):
\begin{equation}
\chi_{n} = 1 + \frac{1-\xi^{n}}{\xi} h_{0} + \ldots
\end{equation}
We can find the error in the finite-volume susceptibility as an
approximation to $\chi$.  For large $n$ (small $\tilde{y}_{n}$), we get
$\chi_n \simeq \chi (1 - \tilde{y}_{n}/\xi) =
 \chi (1 - \xi^{n-1} \tilde{y}_{0})$.
The relative error is then:
\begin{equation}
\frac{\chi-\chi_{n}}{\chi} \simeq \xi^{n-1} \tilde{y}_{0} 
= \xi^{n-1} h_{0} \chi \ .
\label{eq:error}
\end{equation}
This behavior is observed 
\cite{finite} in the approach to infinite volume for the HM.

\subsection{Expansion near the critical point}

In addition of using the HT expansion (expressing $\chi$ in terms of $h_{0}$),
we would also like to have an expansion
in terms of the dual variable $d_{0}$.
Multiplying the numerator and the denominator of 
Eq. (\ref{eq:suscep}) by $y_0^{\gamma}$, we obtain
\begin{equation}
\chi = \frac{\Theta}{(1 - d_{0}) y(d_{0})^{\gamma}} \ .
\end{equation}
with
\begin{equation}
\Theta\equiv\tilde{y}_{0} y_{0}^{\gamma}\ .
\end{equation}
But since $\lambda^{\gamma} = \xi^{-1}$, we have also 
\begin{equation}
\Theta =\tilde{y}_{n} y_{n}^{\gamma}\ ,
\end{equation}
for any $n$. In other words, $\Theta$ is RG invariant.
Due to the discrete nature of our RG transformation, $\Theta$ 
is not exactly a constant. If expressed in terms of ln($y_0$),
$\Theta$ is a periodic function of period ln$\lambda$.
However, for $\lambda$ not too close to 2, the non-zero Fourier modes
are very small \cite{dual}.

\subsection{Numerical evidence for overlapping domains of convergence}

As we have seen above, $\tilde{y}_{n} y_{n}^{\gamma} $ is the same for 
every $n$. We can thus pick $n$ such that we are just in the crossover
region and $both$ expansions have a reasonable chance to be accurate.
In order to test the accuracy of the approximations $y_{app}(d)$
(series expansion up to a certain order) for various $n$,
we have prepared an empirical sequences of $d_n$ starting with 
$d_0=10^{-8}$. We have then tested the scaling properties
by calculating
\begin{equation}
D_n=|\left[y_{app}(d_n)/(y_0\lambda^n)\right]-1|\ ,
\label{eq:bdn}
\end{equation}
where $y_0$ was calculated with 16 digits of accuracy.
Optimal approximations are those for which $D_n\simeq 10^{-16}$.
For such approximation, scaling is as good as it can possibly be.
Indeed, due to the peculiar way numerical errors propagate \cite{numerr},
one does not reach exactly the expected level $10^{-16}$ (more about 
this question in section \ref{sec:hmseries}).
We can define a similar dual quantity by replacing $d$ by $h$ and
$y$ by $\tilde{y}$. In this case, $\tilde{y}_0$ is estimated with 
the same accuracy as $y_0$ by stabilizing $\tilde{y}(h_n)/\xi^n$,
for large enough $n$. 

\begin{figure}
\vskip0.2in
\centerline{\psfig{figure=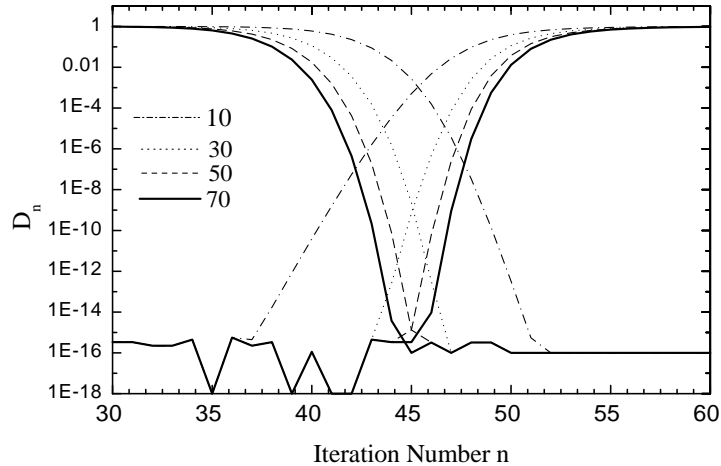,width=4in}}
\caption{Departure from scaling $D_n$ defined in the text, for $y$ (curves
reaching 1 to the right) and $\tilde{y}$ (curves reaching 1 to the left).
In each cases, we have considered approximations of order 10 (dot-dashes),
30 (dots), 50 (dashes) and 70 (solid line). The value of $\lambda$ is 1.5. }
\label{fig:tmoverlap}
\end{figure}

We have performed  this calculation for $\lambda=$ 1.1, 1.5 and 1.9.
The conclusions in the three cases are identical.
For $n$ large enough, the $D_n$ of $y$ starts increasing from $10^{-16}$
until it saturates around 1. 
By increasing the number of terms in the expansion, we can increase
the value of $n$ for which we start losing 
accuracy. Similarly, for $n$ low enough, the $D_n$ of $\tilde{y}$
starts increasing etc... We want to know if it is possible to 
calculate enough terms in each expansion to get scaling with some 
desired accuracy for both functions.
The answer to this question is affirmative 
according to Fig. \ref{fig:tmoverlap} for 
$\lambda =1.5$.

One sees, for instance, that with 10 terms in each series, we have 
scaling with about 1 part in 1000 near $n=45$ for both expansions.
The situation can be improved. For 70-70 expansions, an optimal accuracy
is reached from $n=44$ to 46. For the other values of $\lambda $ quoted
above, similar conclusions are reached, the only difference being
the optimal values of $n$.

\subsection{Analysis of the RG invariant}

\begin{figure}
\centerline{\psfig{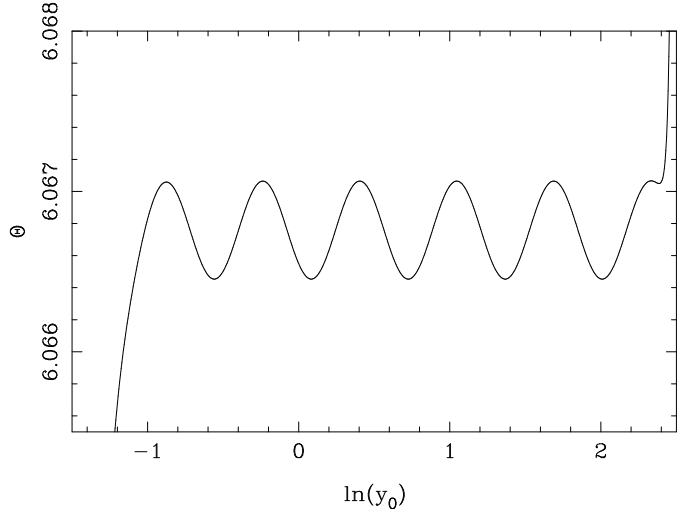}}
\caption{The invariant function $\Theta$, calculated at $\xi = 0.1$,
and plotted against $\ln(y_{0})$.}
\label{fig:all}
\end{figure}

Another evidence for overlapping convergence is that we can calculate
the RG invariant $\Theta$ for a certain range of $y_n$.
To evaluate $\Theta$, we use the series expansions for
$\tilde{y}$ and $d$, cutting each off at some order:
\begin{equation}
\tilde{y}(1 - d(y)) \simeq \sum_{i=1}^{\tilde{m}}
t_{i} (1 - \sum_{j=1}^{m} s_{j} y^j)^i \ ,
\end{equation}
where $s_{l}$ and $t_{l}$ are the $l$th coefficients in the $d$ and
$\tilde{y}$ series, respectively.
The $d$ series is more
accurate the smaller is the value of $y$, while the $\tilde{y}$ series is
more accurate for $d(y)$ close to $1$, which means large values of $y$.
Thus for $y$ very large, we need many terms in the $d$ series, while
we need less terms in the $\tilde{y}$ series.
On the other hand, when $y$ is very small, we need a large value
for $\tilde{m}$ and a relatively small $m$.
We have found that, given a fixed value of $m + \tilde{m}$, the most
accurate values for $\Theta$ are obtained when $m \simeq \tilde{m}$.
In Fig.~\ref{fig:all}, we show
$\Theta$ calculated by keeping $50$ terms
each in the expansions for $y$ and $\tilde{y}$.  The result is plotted
against $\ln(y)$. We used $\xi = 0.1$, which makes the oscillations
much larger than, for example, near to $\xi = 2^{-2/3}$.
Near the fixed points, we need more terms in the appropriate series
to get accurate results.

We can study the oscillation we see in $\Theta$ by looking at
its Fourier expansion.
Since $\Theta$ is periodic in $\ln y_{0}$, we can express
\begin{equation}
\Theta (y_0) = y_0^{\gamma}
\tilde{y}(1-d(y_{0})) = \sum_{p} a_{p} e^{i p \omega \ln y_{0}} \ ,
\end{equation}
where $\omega \equiv 2 \pi/\ln \lambda$.
The coefficients are given as
\begin{equation}
a_{p} = \frac{1}{\ln \lambda} \int_{y_{a}}^{\lambda y_{a}}
 y^{\gamma-1-ip \omega} \tilde{y}(1 - d(y)) dy \ ,
\label{eq:fourier}\end{equation}
for some $y_a$ of our choice.

\begin{figure}
\centerline{\psfig{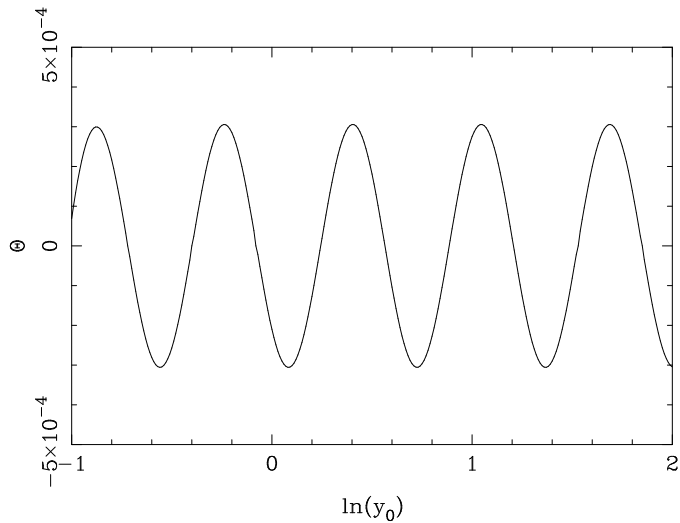}}
\caption{Same as previous plot,
 with the constant term from the Fourier expansion subtracted out.}
\label{allm0}
\end{figure}

\begin{figure}
\centerline{\psfig{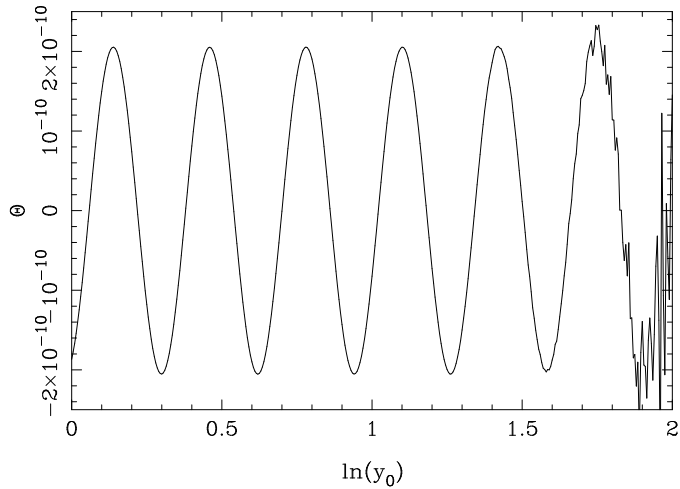}}
\caption{Same as previous plot, with first oscillating term from the
Fourier expansion subtracted out.}
\label{allm0m1}
\end{figure}

As an example, we calculate $a_{0}$ for $\xi = 0.1$ where the oscillations 
are substantial. The choice of the interval can be infered from 
Fig. \ref{fig:all}. 
If we had infinite series, the function would be exactly periodic.
For finite series, we see that $y_a$ cannot be too large or too small.
For intermediate values, we 
obtain $a_{0} \simeq 6.06676$.
In Fig.~\ref{allm0}, we show this constant term subtracted
from the $\Theta$ we evaluated above.  What remains is oscillations
about $0$.  We also calculated $a_{1}$ (using $y_{a} = \lambda$ so
that we keep information about the phase constant as
well as the amplitude).  We construct the first oscillating mode
from this, and in Fig.~(\ref{allm0m1}) we show the result of subtracting out
this in addition to the constant term.  We are left with even smaller
oscillations, from the $a_{2}$ term.

\noindent
In Fig. \ref{allm0m1}, one notices that when 
$\ln (y_0)$ is of the order of 1.7, the breakdown of 
the approximation starts with small numerical noise before 
the complete breakdown near $\ln (y_0)\simeq 2.2$ is 
observed (see Fig. \ref{fig:all}). 

\section{SCALING VARIABLES IN THE HIERARCHICAL MODEL}
\label{sec:multi}

In the one-variable
model, we learned that we could express the basic mapping parameter,
$h_{n}$, the distance from the HT fixed
point, in terms of either of two new variables, $y_{n}$ or $\tilde{y}_{n}$.
The first of these variables transforms multiplicatively with
$\lambda$, the eigenvalue near the unstable fixed point, while
the second transforms  multiplicatively with
$\xi$, the eigenvalue near the HT fixed point.

We would like to extend these methods
to Dyson's HM.
We will express the parameters of the model in terms of
two new sets of variables $y_{n,1},\ldots,
y_{n,l_{max}}$ and $\tilde{y}_{n,1},\ldots,\tilde{y}_{n,l_{max}}$,
 where $l_{max}$ is equal to the number of parameters
we keep in the mapping.  These are the scaling variables.

Near to the unstable fixed point we find a set of
eigenvectors and eigenvalues which control the RG flows
in the linearized region.  We use the notation $\lambda_{i}$
for the $i$th eigenvalue, ordered in size from greatest to least.
For each of these eigenvalues, we introduce a variable $y_{n,i}$ which
transforms under the RG transformation as $y_{n+1,i} = \lambda_{i} y_{n,i}$.
We denote $\tilde{\lambda}_{i}$ for the eigenvalues in the
 linear HT region, similarly defining variables which transform as
$\tilde{y}_{n+1,i} = \tilde{\lambda}_{i} \tilde{y}_{n,i}$.
Then we can expand each of the parameters
in terms of each set of scaling variables.
Our first task will be to use special coordinates where the 
problem will be solved at the linear level. 
To use a relativistic analogy, the transformations leading to
such a system of coordinates (the ``linear scaling variables''), 
can be interpreted as a ``translation'' or shift and a ``rotation''
or more generally a linear transformation.

\subsection{Transformation properties under translation}

We previously considered truncation in the $a_{n,l}$ parameters.
Unfortunately, this approximation becomes less accurate if 
we are interested in the critical region.  We would like to
investigate a truncation that works near to the nontrivial
fixed point.  We re-express the basic recursion in terms
of a new set of parameters, which give the distance from
the nontrivial fixed point. For notational convenience, we 
first rewrite the unnormalized recursion given in Eq. (\ref{eq:truncation2}),
using the ``structure constants'':
\begin{equation}
u_{n,\sigma} = \Gamma_{\sigma}^{ \mu \nu} a_{n,\mu} a_{n,\nu} \ ,
\end{equation}
with
\begin{equation}
\Gamma_{\sigma}^{ \mu \nu}
= \left\{ 
\begin{array}
{r@{\quad,\quad}l}
(c/4)^{\mu+\nu}\ 
\frac{(-1/2)^{\mu + \nu - \sigma}(2(\mu+\nu))!}{ 
(\mu+\nu-\sigma)!(2 \sigma)!}  \ &\   {\rm for}\  \mu+\nu \geq\sigma \\
0\ &\ {\rm otherwise}\ .
\end{array} \right.
\label{eq:struct}
\end{equation}
These zeroes can be understood as a ``selection rule'' 
associated with the fact that $a_{n,l}$ is of order $\beta^l$ 
as explained in section \ref{sec:hm}.
If we follow the truncation procedure explained in section \ref{sec:hm},
the indices simply run over a finite number of values.
We use ``relativistic'' notations. 
Repeated indices mean summation.
The greek indices indices $\mu$ and $\nu$
go from $0$ to $l_{max}$, while latin indices $i$, $j$ go from 1 to   
 $l_{max}$. Obviously,
$\Gamma_{\sigma}^{\mu \nu}$ is symmetric in
$\mu$ and $\nu$.
By construction, $a_{n,0} = 1$ and we can write the normalized 
recursion in the form:
\begin{equation}
a_{n+1,l} = \frac{{\cal M}_l^i a_{n,i}
 + \Gamma_{l}^{ i j} a_{n,i} a_{n,j}}
{1 + {\cal M}_0^i a_{n,i} + \Gamma_{0}^{i j} 
a_{n,i} a_{n,j}} \ ,
\label{eq:general}
\end{equation}
with
\begin{equation}
{\cal M}_{\eta}^{ i} = 2 \Gamma_{\eta}^{ 0 i} \ .
\end{equation}

Let assume that we know a fixed point of the recursion formula
$a_{n,l} = a_{l}^{*}$. We then introduce intermediate variables $g_{n,l}$
which are zero at the fixed point:
\begin{equation}
a_{n,l} \equiv a_{l}^{*} +g_{n,l}\ .
\end{equation}
Plugging this relation into Eq. (\ref{eq:general}), using the fixed point 
properties, subtracting the fixed point and reducing to the same denominator, 
we obtain a recursion formula for the $g_{n,l}$ having the same form as
Eq. (\ref{eq:general}) with the substitutions

\begin{equation}
\Gamma_{\eta}^{ i j} \rightarrow \frac{1}{u_{0}^{*}} 
( \Gamma_{\eta}^{ i j}
 - (1-\delta_{\eta 0}) a_{\eta}^{*} \Gamma_{0}^{ i j}) \ ,
\end{equation}
and
\begin{equation}
{\cal M}_{\eta}^{ i} 
\rightarrow \frac{1}{u_{0}^{*}} 
({\cal M}_{\eta}^{i}+2  \Gamma_{0}^{ij}
 a_{j}^{*}+ 
(1-\delta_{\eta 0})({\cal M}_0^i+2 \Gamma_{0}^{ij} a_{j}^{*})a_{\eta}^{*} \ ,
\end{equation}
where $u_{0}^{*} \equiv \Gamma_{0}^{ \mu \nu} a_{\mu}^{*} a_{\nu}^{*}$.

These equations express the transformation properties of the structure
constants under a translation (shift) 
of the coordinates. In particular if $a_l^{*}=0$
(as for the HT fixed point),
one can see that 
since $\Gamma_{00}^0=1$, the transformation reduces to the identity.
We will now study the transformation properties under changes
of coordinates which will diagonalize the linear RG transformation
matrix ${\cal M}_i^j$.

\subsection{Diagonalization of the linear RG}

The diagonalization of the linear ${\cal M}_i^j$ near the HT fixed 
point is quite simple because it is of the upper triangular form.
The eigenvalues
are just the diagonal terms.  From Eq. (\ref{eq:struct}), 
one sees that $l$th diagonal term is
given as 
\begin{equation}
\tilde{\lambda}_{(r)}=2(c/4)^{r} \ .
\label{eq:hteigenv}
\end{equation}
This spectrum was obtained in Ref. \cite{collet78} with a different method.
This means that an $l \times l$
truncated version of the matrix will have eigenvalues which are identical
to the first $l$ eigenvalues of the full HM map.  Furthermore, the
$l$th eigenvector will contain only $l$ non-zero entries.  This means that
if we truncate to $a_{l_{max}}$, we are simultaneously truncating to
the parameter space to a subspace spanned by the first $l_{max}$ eigenvectors.
This ``stability'' is due to the special relationship existing between 
the HT expansion and the polynomial truncation explained in
section \ref{sec:hm}.

On the other hand, the linearized map near the nontrivial fixed point
does not have these simple properties.
The eigenvalues
may be determined numerically from a truncated version of
the linearized matrix.  With a large enough truncation,
we can find a certain number of the eigenvalues to any desired precision.
We find that there is one eigenvalue larger than $1$, with all the
rest less than $1$.  Though we do not have a simple formula for these
eigenvalues, we know that they do shrink in size quickly,
similarly to the eigenvalues of the HT fixed point. 
For instance the 
numerical values for $c=2^{1/3}$ are $\lambda_1=1.42717\dots$, 
$\lambda_2=0.85941\dots$. A more complete list 
is given in Ref. \cite{gam3}.

Since ${\cal M}$ is not a symmetric matrix the left and 
right eigenvector are distinct.
\begin{equation}
{\cal M}_{l}^{i} \psi^r_i = \lambda_{(r)} \psi^r_l
\end{equation}
and
\begin{equation}
\phi_{r}^{ l} {\cal M}_{l}^{ i} = \lambda_{(r)} \phi_{r}^{ i} \ .
\end{equation}
The notation $(r)$ means that there is no sum on $r$.
Since all the eigenvalues are distinct, one can normalize the
eigenvectors in such a way that
\begin{equation}
\phi_{r}^{ i} \psi^{r^{'}}_ i = \delta_{r}^{ r^{'}} \ . 
\end{equation}
Similarly a completeness relation (decomposition of the identity) 
can be obtained by summing 
over $r$:
\begin{equation}
\phi_{r}^{ i} \psi^{r}_j = \delta_{j}^{i} \ . 
\end{equation}
In numerical calculations, the relations of orthogonality and completeness
provide a  reliable way to check the accuracy of the calculations.

We can thus diagonalize the linearized RG:
\begin{equation}
\phi_{r}^{ l} {\cal M}_{l}^{ i} \psi^{r^{'}}_ i = \lambda_{(r)} 
\delta_{r}^{ r^{'}} \ . 
\end{equation}
If we start with the transformation expressed in the general form
Eq. (\ref{eq:general}) but with $a$ replaced by $g$ to indicate that 
the fixed point is at the origin, as in the previous subsection, we can 
define a new set of coordinates $d_{n,l}$, such that
\begin{equation}
g_{n,l} = \psi^{r}_{ l}d_{n,r} \ ,
\label{eq:transf}
\end{equation}
or, equivalently,
\begin{equation}
d_{n,r} = \phi_{r}^{ l} g_{n,l} \ .
\end{equation}
With these notations, all the structure constants transform according 
to the familiar rules of tensor analysis. 
Since $\phi$ and $\psi$ are inverse of each other, we can think
of lower indices as covariant and upper indices as contravariant.
As an example of transformation under Eq. (\ref{eq:transf}), we have
\begin{equation}
\Gamma_p^{qr}\rightarrow \phi_p^{p^{'}}\Gamma_{p^{'}}^{q^{'}r^{'}}
\psi^q_{q^{'}}\psi^r_{r^{'}}\ .
\end{equation}
The other structure constants transform according to the same rules.

The eigenvectors are not determined uniquely even after normalizing.
If $\psi^r_i$ and $\phi_r ^i$ are
components of normalized right and left eigenvectors, then
$\psi^r_i = \alpha_{(r)} \psi^r_i$ and
$\phi_{r}^i = \phi_{r}^i/\alpha_{(r)}$, where $\alpha_{(r)}$ is
a constant, will work equally well.
In the following, we will fix this ambiguity by requiring that 
the ``other'' fixed point be
at $(1,1,\ldots)$ in the new coordinates.

\subsection{The Canonical Form}

In summary, we can choose a system \
of coordinate $d_l$ where 
the unstable fixed point will be at the origin of the coordinate
and the 
HT fixed point at $(1,1,\dots)$, and such that the linear
RG transformation is diagonal. In this system
of coordinates, the RG transformation reads
\begin{equation}
d_{n+1,r} = \frac{\lambda_{(r)} d_{n,r}
 + \Delta_{r}^{ p q} d_{n,p} d_{n,q} }
{1 + \Lambda^{p} d_{n,p} + \Delta_{0}^{ p q} d_{n,p} d_{n,q} } \ ,
\label{eq:hmdrules}
\end{equation}
where the new structure constant are calculated 
from the original ones according to the 
transformation laws discussed in the previous two subsections.

Similarly,
we can introduce new coordinates $h_{l}$, so that
\begin{equation}
a_{n,l} = \tilde{\psi}^r_l h_{n,r} \ ,
\label{eq:atoh}
\end{equation}
and take the eigenvectors so that the unstable  fixed point is
at $(1,1, \ldots)$ in the new system of coordinates.
Note that the form of the eigenvectors guarantees that 
$h_{n,l}$ is of order $\beta^l$. This can be seen by inverting 
Eq. (\ref{eq:atoh}) using the matrix of left eigenvalues.
Due to the upper-diagonal form, the second left eigenvector has 
its first entry equal to zero, the third its first two entries etc... .
In this new system, the RG transformation reads 
\begin{equation}
h_{n+1,r} = \frac{\tilde{\lambda}_{(r)} h_{n,r}
 + \tilde{\Delta}_{r}^{ p q} h_{n,p} h_{n,q} }
{1 + \tilde{\Lambda}^{p} h_{n,p} 
+ \tilde{\Delta}_{0}^{ p q} h_{n,p} h_{n,q}}\ .
\end{equation}

\subsection{Expansions in terms of the scaling variables}

We would like to express 
the evolution of the canonical coordinates in terms of 
functions of the scaling variables:
\begin{equation}
d_{n,r} = \sum_{i_{1},i_{2},\ldots} s_{r,i_{1} i_{2} \ldots} y_{n,1}^{i_{1}}
y_{n,2}^{i_{2}} \ldots \ ,
\end{equation}
where the sums over the $i$'s run from $0$ to infinity in each variable
and $y_{n,l}=\lambda_{(l)}^n y_{0,l}$.
In practice, we expand in terms of some finite number of the scaling
variables, less than or equal to the number of $d$ parameters we
have truncated to.  Using the notation ${\mathbf{i}} = (i_{1},i_{2} \ldots$)
and the product symbol, we may rewrite the expansion as
\begin{equation}
d_{n,r} = \sum_{\mathbf{i}} s_{r,\mathbf{i}} \prod_{m} y_{n,m}^{i_{m}}
\end{equation}
Using the transformation law for the scaling variables, we have
\begin{equation}
d_{n+1,r} = \sum_{\mathbf{i}} s_{r,\mathbf{i}}  \prod_{m} (\lambda_{(m)}
 y_{m})^{i_{m}} \ .
\end{equation}
Each constant term, $s_{r,0,0,\ldots}$, is zero, as the scaling
variables vanish at the fixed point.  From Eq.~(\ref{eq:hmdrules}),
we see that all but one of the linear terms are zero for each value of $r$.  
The remaining term is the one proportional to the $r$th scaling variable.
We take these coefficients to be $1$, so that the $d_{n,r} \simeq y_{n,r}$
for small $y_{n,r}$.  For the higher-order terms, we obtain the recursion
\begin{equation}
s_{r, \mathbf{i}} = \frac{ \sum_{\mathbf{j}+\mathbf{k} = \mathbf{i}}
 ( \Delta_{r}^{ p q} s_{p,\mathbf{j}}
 s_{q,\mathbf{k}} - s_{r,\mathbf{j}} \prod_{m} \lambda_{(m)}^{j_{m}}
\Lambda^{p} s_{p, \mathbf{k}} ) -\sum_{\mathbf{j}
+\mathbf{k}+\mathbf{l}=\mathbf{i}}
 s_{r, \mathbf{j}} \prod_{m} \lambda_{(m)}^{j_{m}} \Delta_{0}^{ p q}
 s_{p, \mathbf{k}} s_{q, \mathbf{l}} }
{\left(\prod_{m} \lambda_{(m)}^{i_{m}} - \lambda_{r}\right)} \ .
\label{eq:sca}
\end{equation}
The calculation can be organized in such way that the r. h. s. of the 
equations are already known. This will be the case for instance if 
we proceed order by order in $\sum_q i_q$, the degree of non-linearity.
As one can see, this expansion may suffer from ``small
denominator problems''. This issue will be discussed elsewhere 
\cite{ymprogress}.

We can likewise expand each $h_{n,r}$ in terms of scaling variables
$\tilde{y}_{n,1},\tilde{y}_{n,2}, \ldots$.  The derived recursions
are identical in form to those derived above. From Eq. (\ref{eq:hteigenv}),
one sees that the denominator will vanish for some equations and 
unless the numerator is also zero, the expansion is ill-defined.
Again this question will be discussed elsewhere \cite{ymprogress}.
In the following, we will only use the expansion for the $d$-functions.

\subsection{Expansion of the scaling variables }

One can likewise find expansions of the scaling variables in terms of 
the canonical coordinates, by setting 
\begin{equation}
y_{n,r} = \sum_{\mathbf{i}} u_{r,\mathbf{i}} \prod_{m} d_{n,m}^{i_{m}}\ ,
\end{equation}
and requiring that when ${\bf d}_n$ is replaced by ${\bf d}_{n+1}$, the 
function is multiplied by $\lambda _{(r)}$. Since ${\bf d}_{n+1}$ has a
denominator, it needs to be expanded for instance in increasing order
of non-linearity. The same considerations applies for $\tilde{y}$.
Note that small denominators may also be present in these calculations. 
However, the calculation of $y_1$ and $\tilde{y}_1$ is free of such a problem
since the largest eigenvalue cannot be written as product of lower eigenvalues 
smaller than 1.

It is in principle simple to obtain these expansions
order by order in the degree of non-linearity, however
there exists some practical limitations. For instance, if we want to
calculate all the non-linear terms of order 10 in any of the
variables, with $l_{max}$=30, we need 
to calculate and store $30^{10}\sim 10^{15}$ numbers.
In the next section, we show that one can organize these calculations
in a way which allows accurate answers for the susceptibility.

\subsection{The susceptibility}

From Eqs. (\ref{eq:sus}) and (\ref{eq:atoh}), we obtain
\begin{equation}
\chi_{n} = -(2/\beta)\tilde{\psi}^r_1 h_{n,r} (2/c)^n \ .
\end{equation}
For $n$ large enough, the linear behavior applies and the the 
$h_{n,r}$ get multiplied by $2(c/4)^r$ at each iteration.
In the large $n$ limit, only the $r=1$ term survives and 
consequently,
\begin{equation}
\chi=  -(2/\beta)\tilde{\psi}^1_1 {\rm lim}_{n\rightarrow\infty}
h_{n,1} (2/c)^n \ .
\end{equation}
Using the same method as in the one-variable model, we 
can in the limit replace $h_{n,1}$ by $\tilde{y}_{n,1}$ and obtain
\begin{equation}
\chi=  -(2/\beta)\tilde{\psi}^1_1 \tilde{y}_{0,1}\  .
\end{equation}
This allows us to write
\begin{equation}
\chi=  -(2/\beta)\tilde{\psi}^1_1 \Theta\tilde{y}_{0,1}^{-\gamma}\  ,
\end{equation}
where $\gamma \equiv -\ln(\tilde{\lambda}_{1})/\ln(\lambda_{1})$
and $\Theta$ the RG invariant
\begin{equation}
\Theta \equiv \tilde{y}_{0,1} y_{0,1}^{\gamma} =
 \tilde{y}_{n,1} y_{n,1}^{\gamma} \ .
\label{eq:susyy}
\end{equation}
For reference, in the case $c=2^{1/3}$, $\tilde{\psi}^1_1
\simeq -0.564$.

One can calculate the subtracted four-point function following the same 
procedure, namely expressing the $a_{n,l}$ in terms of the 
$h_{n,l}$. However, $\tilde{\lambda}_1^2>\tilde{\lambda}_2$ and one needs to 
go beyond the linear expansion to calculate these quantities. 

\section{Approximations and numerical implementation}
\label{sec:impl}

In this section, we show that it is possible to design approximations
such that one can calculate the susceptibility using Eq. (\ref{eq:susyy}).
For this purpose we have calculated an empirical series of $a_{n,l}$ with
$c=2^{1/3}$, $\beta=\beta_c-10^{-8}$ and an initial Ising measure. 
Detail relevant 
for this calculations can be found in Refs. \cite{finite,gam3,numerr}.
The calculations have been performed with $l_{max}=30$, a value for
which at the $\beta$ considered, the errors due to the 
truncation are of the same order as those due to the numerical errors.
These errors are small enough to 
allow a determination of the susceptibility with seven
significant digits if we use double precision.

The empirical flow
proceeds in two steps. First, the flow goes from the
initial mesure to close to the unstable fixed point.
Second, the flow goes from close to the unstable fixed point to 
the HT fixed point. 
The first step depends on the choice of the initial measure and
will not 
be discussed in full detail. Our main goal will be to show that it is
possible to construct nonlinear expansions which allow to describe 
accurately the second step.
We now discuss the flow chronologically. In order to keep track of 
the chronological sequence, we postponed the 
technical discussion of the
convergence of the series to section \ref{sec:hmseries}. In this section,
the results are simply quoted.

Our choice of $\beta$ means that we start near the stable manifold.
After about 25 iterations, we start approaching the unstable fixed point
and the linear behavior $d_{n+1,l}\simeq\lambda_l d_n$ becomes a good
approximation. During the next 20 iterations, the irrelevant variables
die off at the linear rate and at the same time we move away from the 
fixed point along the unstable direction, 
also at the linear rate. At $n=47$ we are in good approximation on the 
unstable manifold and $d_{n,2}$ becomes proportional to $d_{n,1}^2$.
In other words, the non-linear terms are taking over.
At this point, we can approximate the $d_{n,l}$ as functions
of $y_1$ only: $d_{n,l}\simeq d_l(\lambda_1^n y_1,0,0,\dots)$. This 
approximation is consistent in the sense that if $y_2=0$ at $n=0$
then it is also the case for positive $n$. 

We have calculated $d_l(y_1,0,0,\dots)$ up to order 80 in $y_1$ 
using Eq. (\ref{eq:sca}). We have then inverted $d_1$ to obtain $y_1$.
Given the empirical $d_{n,1}$ we then calculated the approximate
$y_{n,1}$ and then used the other functions $d_l(y_1)$ 
(with $l\geq2$) to ``predict''
$d_{n,l}$. Comparison with the actual numbers were good in a restricted 
range. For $n=49$, the relative errors were
less than a percent. They kept decreasing to less than one part in 
10,000 for 
$n=54$ and then increased again. It will be shown in section
\ref{sec:hmseries} that this corresponds to the fact that
when $y_1$ becomes too large (a value of 3.7 first exceeded at
$n=57$), the series expansion of $d_1$ diverges, unlike the 
one-variable model for which $d(y)$ is analytical.
It will also be shown later that the quality of the approximation between 
$n=45$ and $n=55$ can be improved by  
by treating $y_2$, $y_3,\dots$ perturbatively. 

Near $n=55$, the presence of the HT fixed point starts dominating
the flow but we are still far away from the linear regime. To take into 
account the non-linear effects, we have calculated $\tilde{y}_1({\bf h})$
up to order 11 in $\beta$. The reason for this modest order is that this 
is a multi-variable expansion. Recalling the discussion about the HT
expansion in section \ref{sec:hm}, we can count the number of terms at each
order in $\beta$. At order two, we have $h_1^2$ and $h_2$, but since 
the linear transformation is diagonalized, $h_2$ will only appear in 
$\tilde{y}_2({\bf h})$. At order three, we have $h_1^3$, $h_2h_1$ and not
$h_3$. It is easy to see that at order $l$, one has $p(l)-1$ terms, where
$p(l)$ is the number of partitions of $l$. It has been known from the 
work of Hardy and Ramanujan that
\begin{equation}
p(l)\sim \frac{{\rm exp}(\pi \sqrt{2l/3})}{4\sqrt{3} l}\ .
\end{equation}
It seems thus difficult to get very high order in this expansion.
In order to fix the ideas, there are 41 terms at order 11, 
489 at order 20 and 13,848,649 at order 80.

As explained in section \ref{sec:hmseries},
the expansion up to order 11 has a sufficient accuracy to take over
at $n=55$. It also provides
optimal (given our use of double precision) results for $n\geq 60$.
As $n$ increases beyond 60, one can see the effect of each order
disappear one after the other. Finally, the linear behavior becomes
optimal near $n=130$. 
This concludes our chronological discussion. The main point made was
that there exists a region in the crossover where both expansions
are valid. We now proceed to justify empirically the claims made above.

\section{Large-order behavior for the HM}
\label{sec:hmseries}

\subsection{$y_1(d_1)$}

\begin{figure}
\vskip0.2in
\centerline{\psfig{figure=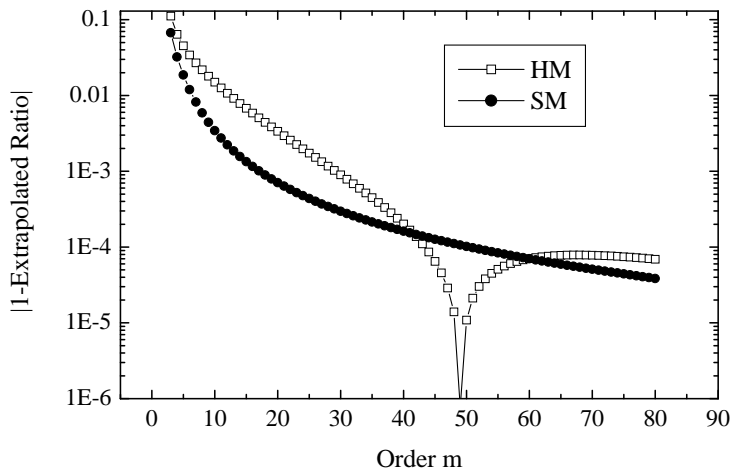,width=4.0in}}
\caption{The absolute value of the difference between the extrapolated ratio
and 1 for the HM (empty boxes) and the SM (full circles),
as a function of the order.}
\label{fig:rhdone}
\end{figure}

\begin{figure}
\vskip0.2in
\centerline{\psfig{figure=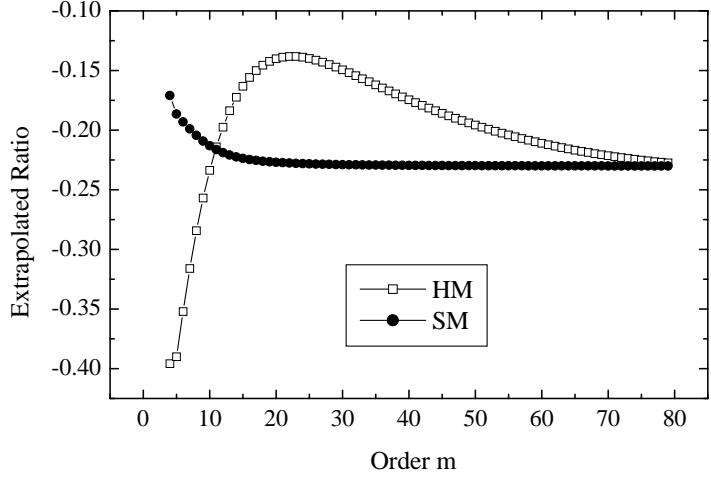,width=4.0in}}
\caption{The extrapolated slope $\hat{S}_m$ for the HM (empty boxes) 
and the SM (full circles) as a function of the order .}
\label{fig:shdone}
\end{figure}

The behavior of $y_1(d_1)$ obtained by the procedure described above,
has been studied following the same methods as for the one-variable 
model. In order to provide a comparison, we have calculated the same
quantities for the one-variable model with $\lambda=1.2573$. 
In the following, we call this model the ``simplified model'' (SM).
For this
special value, the critical exponents $\gamma$ of the two models
coincide with five significant digits. 
We have good evidence that both series have a radius 
of convergence 1 as indicated by the extrapolated ratio 
defined in Eq. (\ref{eq:rhat}) reaching
1 at an expected rate (Fig. \ref{fig:rhdone}).
Similarly, their extrapolated slopes seem to converge to the same value 
$1/\gamma -1\simeq -0.23026\dots $ as shown in Fig. \ref{fig:shdone}.
In conclusion, the function $y(d)$ for the SM is a 
reasonably good model to 
guess the asymptotic behavior of $y_1(d_1)$.

\subsection{$d_1(y_1)$}

\begin{figure}
\vskip0.2in
\centerline{\psfig{figure=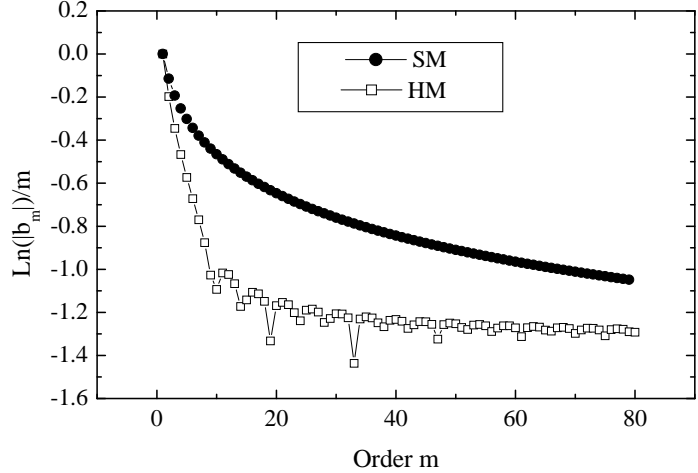,width=4.0in}}
\caption{Logarithm of the absolute value of the coefficients of the 
expansion of $d_1(y_1)$ divided by the order, for the HM (empty boxes) and the 
SM (circles).}
\label{fig:logco}
\end{figure}

The situation is different for the inverse function $d_1(y_1)$.
A first look at the differences is given in Fig. \ref{fig:logco}.
%\begin{figure}
%\vskip0.2in
%\centerline{\psfig{figure=lnrathm.EPS,width=4.0in}}
%\vskip0.2in
%\caption{for the HM (empty boxes) and the SM (full circles }
%\end{figure}
\noindent
The quantity plotted in this figure will be used to discriminate
between a finite and an infinite radius of convergence. If 
$|b_m|\sim R^{-m}$ as for a radius of convergence $R$, then 
we have ${\rm ln}(|b_m|)/m \sim -{\rm ln}(R)+A/m$ for some constant $A$.
On the other hand, if $|b_m|\sim (m !)^{-\alpha}$ as for an infinite 
radius of convergence, then 
we have ${\rm ln}(|b_m|)/m \sim -\alpha({\rm ln}(m)-m)$.
In the following, we will compare fits of the form $A_1+A_2/m$ and
$B_1{\rm ln}(m)+B_2$.

\begin{figure}
\centerline{\psfig{figure=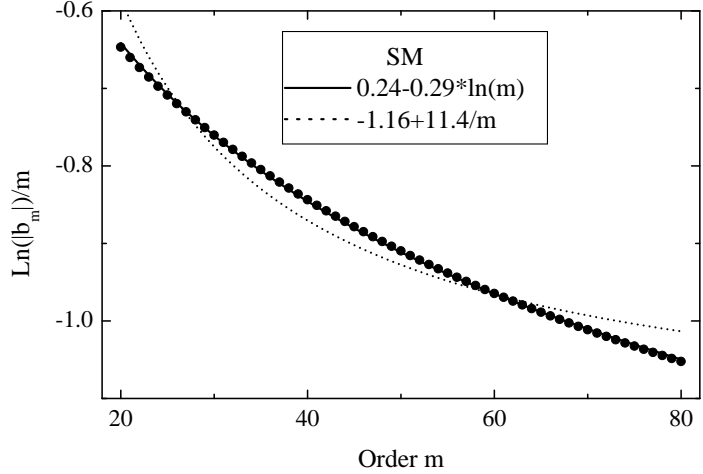,width=4.0in}}
\caption{Comparison of fits of the form $A_1+A_2/m$ (dots)and
$B_1{\rm ln}(m)+B_2$ (solid line) 
with the data provided in Fig. \ref{fig:logco}
for the SM (circles).}
\label{fig:toyfit}
\end{figure}

We first consider the case of SM, where according to the our study in
section \ref{sec:largetm}, we should have an infinite radius of 
convergence. This possibility is highly favored as shown in Fig. 
\ref{fig:toyfit}.
\noindent
One sees clearly that the solid line is a much better fit.
The chi-square for the solid line fit is 200 times smaller.
In addition $B_1\simeq-B_2$ as expected.
In conclusion, this analysis comfirms the ratio analysis done previously
and favors strongly the inifite radius of convergence possibility. 

\begin{figure}
\centerline{\psfig{figure=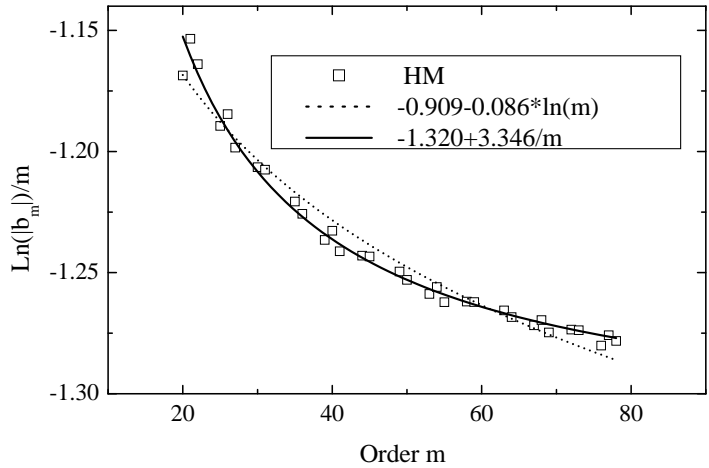,width=4.0in}}
\caption{Comparison of fits of the form $A_1+A_2/m$ (solid line)and
$B_1{\rm ln}(m)+B_2$ (dots) with selected points of the data 
in Fig. \ref{fig:logco}
for the HM  (boxes).}
\label{fig:hmfit}
\end{figure}

The analysis for the HM is more delicate. One observes periodic
``dips'' in Fig. \ref{fig:logco} which make the ratio analysis almost 
impossible. We have thus only considered, the ``upper envelope'' by 
removing the dips from Fig. \ref{fig:logco}. The fit represents an upper 
bound rather than the actual coefficients. The fits of the upper envelope
are shown in Fig. \ref{fig:hmfit}
The possibility of a finite radius of convergence is slightly favored, the 
chi-square being 0.4 of the one for the other possibility. Also, the second 
fit does not have the $B_1\simeq-B_2$ property. From $A_1\simeq-1.32$, we 
estimate that the radius of convergence is about 3.7. 

\subsection{$\tilde{y}_1({\bf h})$}

\begin{figure}
\vskip0.2in
\centerline{\psfig{figure=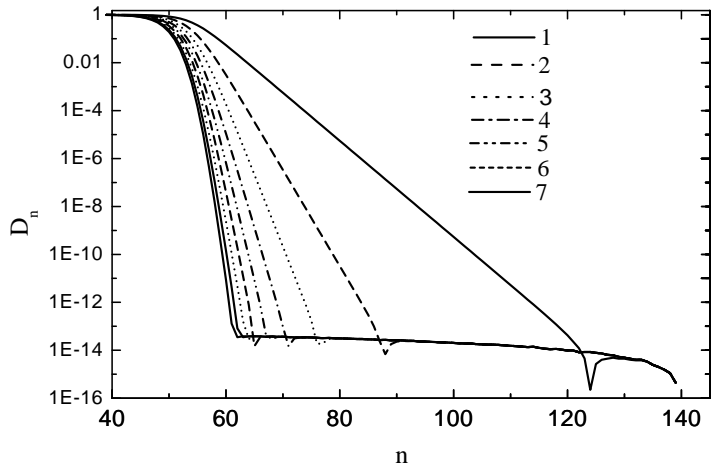,width=4.0in}}
\vskip0.2in
\caption{The quantity $D_n$ defined in Eq. (\ref{eq:bdn}) for expansions 
of $\tilde{y}_1$ in $\beta$, at order 1 (solid line), 2 (dashes), 3 (dots),
etc... for each iteration $n$.}
\label{fig:app}
\end{figure}

As explained in section \ref{sec:impl}, one can calculate 
$\tilde{y}_1({\bf h})$
using an expansion in $\beta$. 
As in section \ref{sec:largetm}, we will use an empirical series
$a_{n,l}$, calculate the corresponding $h_{n,l}$ an plug them 
in the scaling variables. 
This empirical series was calculated with an initial Ising measure and 
$\beta=\beta_c -10^{-8}$ (see Ref. \cite{gam3}).
Again we define a quantity $D_n$ as in Eq.
(\ref{eq:bdn}) which is very small when we have good scaling and increases
when the approximation breaks down. The results are shown in 
Fig. \ref{fig:app} for successive orders in $\beta$.
The solid line on the right is the linear approximation. It becomes optimal
near $n=130$. The next line (dashes) is the second order in $\beta$ expansion.
It becomes optimal near $n=90$. Each next order gets 
closer and closer to be optimal near $n=60$. 
The last curve on the left is the order 9 approximation. It is hard to 
resolve the next two approximations on this graph.

\begin{figure}
\centerline{\psfig{figure=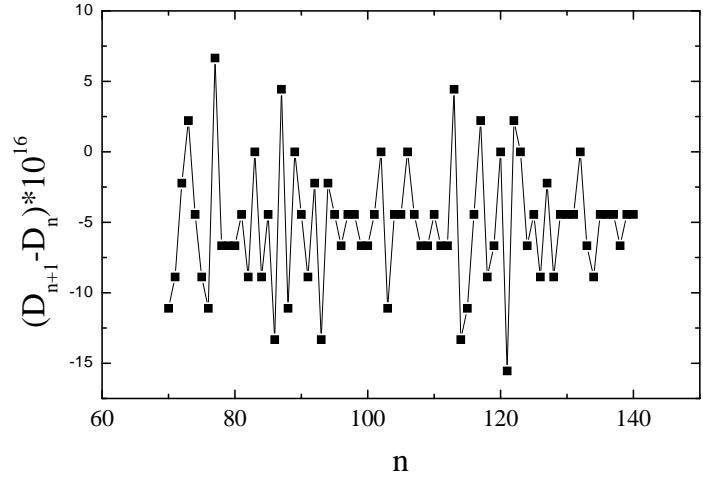,width=4.0in}}
\caption{Difference between two successive $D_n$ in $10^{-16}$ units.}
\label{fig:numerr}
\end{figure}

The asymptotic value is stabilized with 16 digits and one may wonder
why we get only scaling with 14 or 15 digits in Fig. \ref{fig:app}.
The reason is that we use empirical data and that numerical errors can 
add coherently as explained in Ref. \cite{numerr}. This can be seen 
directly by considering the difference between two successive values 
of $D_n$ as shown in Fig. \ref{fig:numerr}.
\noindent
One sees that the numerical errors at each step tend to be negative 
more often than positive, and consequently there is a small ``drift''
which affects the last digits.

\subsection{overlapping domains of convergence}

\begin{figure}
\centerline{\psfig{figure=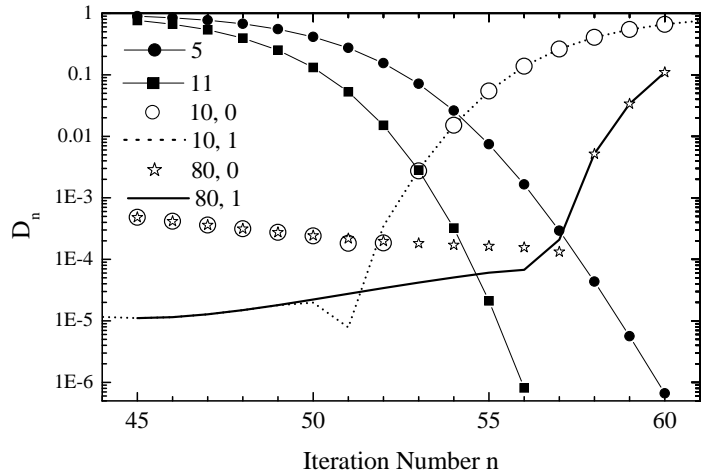,width=4.0in}}
\caption{Values of $D_n$ for $\tilde{y}_1$ up to order 5 (filled circle)
and 11 (filled boxes), and for $y_1$ up to order 10 in $d_1$ (empty circles)
and with first order corrections in $y_2$ (dots), and up to order 80
in $d_1$ (empty stars) and with first order corrections in $y_2$ (solid line).}
\label{fig:hmoverlap}
\end{figure}

We can now look at the $D_n$ defined as in Eq. (\ref{eq:bdn}) 
for $y_1$ and $\tilde{y}_1$ together 
in Fig. \ref{fig:hmoverlap}.
\noindent
If we use an expansion of order 5 in $\beta$ for $\tilde{y}$ 
and of order 10 in $d_1$ for $y_1$, we can get scaling within a few 
percent for both variables at $n=54$. We can go below 1 part in 1000,
with an expansion of order 11 in $\beta$ and order 80 in $d_1$.
At this point, the main problem is that the effects of the subleading 
correction makes the scaling properties {\it worse} when $n\leq 57$
and $n$ decreases. One can improve the scaling properties by taking
the effects of $y_2$ into account. A detailed study shows that one 
can estimate the subleading effects between $n=40$ and $n=45$.
One finds that
\begin{equation}
\frac{y_1(d_{n,1})}{\lambda_1^n}\simeq 7.2778\times 10^{-9}-4.5
\times 10^{-10}\times\lambda_2^n 
\end{equation}
It is thus possible to get a function scaling better by subtracting these
correction. This improve the scaling properties by almost one order of 
magnitude near $n=54$ and by almost two order of magnitude near $n=45$.
It is likely  that our approximation of having a single 
scaling variable can be corrected order by order in $y_2$, $y_3$ etc... .
\section{conclusions}

We have shown in two examples that the susceptibility can be expressed 
in terms of the scaling variables corresponding to the two fixed points
governing the HT flows. 
We have given convincing evidence that the
expansions of these variables have overlapping domains of convergence.
Several interesting questions remain to be discussed.

We have not discussed in detail the initial approach of the 
unstable fixed point. We have just shown that it can in principle 
be incorporated by calculating the 
subleading corrections in $y_2$, $y_3$ etc... This study 
depends on the details of the initial measure. A particularly 
interesting set of initial measures are the Landau-Ginzburg models
in the vicinity of the Gaussian fixed point. We can use
the usual perturbation theory 
in the quartic (or higher orders) 
coupling constant to construct the scaling variables associated 
with the Gaussian fixed point following the procedure described above, and
interpolate between the Gaussian fixed point and the unstable 
fixed point.

The question of small denominators and resonances have been circumvented 
by using only quantities which are free of this problem (namely $y_1$ and
$\tilde{y}_1$ ) in our calculation of the susceptibility. 
The effects of the small denominators of the other scaling
variables is a complex topic presently under study \cite{ymprogress}.

We emphasize that the complete 
knowledge of the scaling functions and their
inverse provides analytical expressions for all the thermodynamic
quantities at any volume (see Eq. (\ref{eq:dsubn}) in the simplified model).
In practical calculations, one is naturally led to combine different expansions
and it is thus important to use sets of interactions which are compatible,
as done in Ref. \cite{taro00} for gauge theories.

\begin{acknowledgments}
This research was supported in part by the Department of Energy
under Contract No. FG02-91ER40664.
Y. M. thanks the 
Aspen Center for Physics for its hospitality in Summer 1999, where discussions
with P. de Forcrand motivated the completion of this work 
and in Summer 2000 while 
this manuscript was written.
\end{acknowledgments}
%\bibliography{da} 

\begin{references}
\bibitem{wilson71a}
K. Wilson, Phys.\ Rev.\ D {\bf 3},  1818  (1971).

\bibitem{wilson74}
K. Wilson and J. Kogut, Phys.\ Rep. \ {\bf 12},  75  (1974).

\bibitem{taro00}
P. de~Forcrand~et al., Nucl. Phys. B {\bf 577},  263  (2000).

\bibitem{gonzalez87}
A. Gonzalez-Arroyo and M. Okawa, Phys. Rev. D {\bf 35},  672  (1987).

\bibitem{dyson69}
F. Dyson, Comm.\ Math.\ Phys.\ {\bf 12},  91  (1969).

\bibitem{baker72}
G. Baker, Phys.\ Rev.\ B {\bf 5},  2622  (1972).

\bibitem{baker77}
G. Baker and G. Golner, Phys. Rev. B {\bf 16},  2080  (1977).

\bibitem{kim77}
D. Kim and C. Thomson, J. Phys. A {\bf 10},  1579  (1977).

\bibitem{finite}
J. Godina, Y. Meurice, M. Oktay, and S. Niermann, Phys. Rev. D {\bf 57},  6326
  (1998).

\bibitem{gam3rapid}
J. Godina, Y. Meurice, and M. Oktay, Phys. Rev. D {\bf 57},  R6581  (1998).

\bibitem{gam3}
J. Godina, Y. Meurice, and M. Oktay, Phys. Rev. D {\bf 59},  096002  (1999).

\bibitem{wegner72}
F. Wegner, Phys. Rev. B {\bf 3},  4529  (1972).

\bibitem{cardy96}
J. Cardy, {\em Scaling and Renormalization in Statistical Physics} (Cambridge
  University Press, Cambridge, 1996).

\bibitem{dual}
Y. Meurice and S. Niermann, Phys. Rev. E {\bf 60},  2612  (1999).

\bibitem{hyper}
J.~J. Godina, Y. Meurice, and M. Oktay, Phys. Rev. D {\bf 61},  114509  (2000).

\bibitem{wittwer95}
H. Koch and P. Wittwer, Math. Phys. Electr. Jour. {\bf 1},  Paper 6  (1995).

\bibitem{nickel80}
B. Nickel,  in {\em Phase Transitions, Cargese 1980}, edited by M. Levy, J.~L.
  Guillou, and J. Zinn-Justin (Plenum Press, New York, 1982).

\bibitem{osc1}
Y. Meurice, G. Ordaz, and V.~G.~J. Rodgers, Phys.\ Rev.\ Lett. {\bf 75},  4555
  (1995).

\bibitem{osc2}
Y. Meurice, S. Niermann, and G. Ordaz, J.\ Stat.\ Phys.\ {\bf 87},  363
  (1997).

\bibitem{collet80}
P. Collet and J.-P. Eckmann, {\em Iterated of the interval as dynamical
  systems} (Birkhauser, Boston, 1980).

\bibitem{arnold88}
V. Arnold, {\em Geometrical Methods in the Theory of Ordinary Differential
  Equations} (Springer-Verlag, New York, 1988).

\bibitem{numerr}
Y. Meurice and B. Oktay, Non-Gaussian numerical errors versus mass hierarchy,
  hep-lat/0005011.

\bibitem{collet78}
P. Collet and J.-P. Eckmann, {\em A Renormalization Group Analysis
 of the Hierarchical Model in Statistical Mechanics}, 
Edited by J. Ehlers et al., Lecture Notes in 
Physics {\bf 74}, (Springer-Verlag, 
Berlin, 1978).


\bibitem{ymprogress}
Y. Meurice, in preparation.
\end{references}
%\bibliographystyle{unsrt}
%\bibliographystyle{prsty}
%\end{document}
%\begin{thebibliography}{10}

\end{document}